\documentclass[aps,prb,reprint,longbibliography,superscriptaddress]{revtex4-2}
\usepackage{graphicx}  
\usepackage{dcolumn}   
\usepackage{tikz}
\usepackage{bm}        
\usepackage{amssymb}   
\usepackage{framed}
\usepackage{amsmath}
\usepackage{appendix}
\usepackage{hhline}
\definecolor{bluegreen}{rgb}{0,0.2,0.8}
\usepackage{subfigure,amsmath,verbatim,moreverb}
\usepackage{tabularx}
\usepackage{adjustbox}
\usepackage{lipsum}
\usepackage{longtable}
\usepackage{booktabs}
\usepackage{latexsym}
\usepackage{dcolumn}
\usepackage{amsmath}
\usepackage{epsf}
\usepackage{float}
\usepackage[breaklinks=true,colorlinks,citecolor=blue,linkcolor=blue,urlcolor=blue]{hyperref}
\usepackage[normalem]{ulem} 
\usepackage{multirow}
\usepackage{siunitx}
\usepackage{longtable}
\usepackage{booktabs}
\usepackage{threeparttable}
\usepackage[T1]{fontenc}
\usepackage{booktabs}
\usepackage{multirow}
\usepackage{graphicx}
\renewcommand{\arraystretch}{1.3} 

\usepackage{color}

\usepackage{etoolbox}
\AtBeginEnvironment{align}{\setcounter{subeqn}{0}}
\newcounter{subeqn} %

\begin{document}

\title{
Advancing excited-state properties of two-dimensional materials using a dielectric-dependent hybrid functional
}
\author{Arghya Ghosh}
\affiliation{Department of Physics, Indian Institute of Technology, Hyderabad, India}
\author{Subrata Jana}
\email{Contact author: subrata.niser@gmail.com, subrata.jana@umk.pl}
\affiliation{Institute of Physics, Faculty of Physics, Astronomy and Informatics, Nicolaus Copernicus University in Toru\'n, ul. Grudzi\k{a}dzka 5, 87-100 Toru\'n, Poland}
\author{Manoar Hossain}
\affiliation{Paderborn Center for Parallel Computing, Paderborn University, Warburger Str. 100, 33098 Paderborn, Germany}
\author{Dimple Rani}
\affiliation{School of Physical Sciences, National Institute of Science Education and Research, An OCC of Homi Bhabha National Institute, Bhubaneswar 752050, India}
\author{Szymon \'Smiga}
\affiliation{Institute of Physics, Faculty of Physics, Astronomy and Informatics, Nicolaus Copernicus University in Toru\'n, ul. Grudzi\k{a}dzka 5, 87-100 Toru\'n, Poland}
\author{Prasanjit Samal}
\affiliation{School of Physical Sciences, National Institute of Science Education and Research, An OCC of Homi Bhabha National Institute, Bhubaneswar 752050, India}

\begin{abstract}
Predicting accurate band gaps and optical properties of lower-dimensional materials, including two-dimensional van der Waals (vdW) materials and their heterostructures, remains a challenge within density functional theory (DFT) due to their unique screening compared to their bulk counterparts. 
Additionally, accurate treatment of the dielectric response is crucial for developing and applying screened-exchange dielectric-dependent range-separated hybrid functionals (SE-DD-RSH) for vdW materials. In this work, we introduce a SE-DD-RSH functional to 
the 2D vdW materials like MoS$_2$, WS$_2$, $h$BN, black phosphorus (BP), and $\beta$-InSe.  
By accounting for in-plane and out-of-plane dielectric responses, our method achieves accuracy comparable to advanced many-body techniques like $G_0W_0$ and BSE@$G_0W_0$ at a 
lower computational cost. 
We demonstrate improved band gap predictions and optical absorption spectra for both bulk and layered structures, including some heterostructures like MoS$_2$/WS$_2$. This approach offers a practical and precise tool for exploring electronic and optical phenomena in 2D materials, paving the way for efficient computational studies of layered systems.

\end{abstract}

\maketitle

\section{Introduction}

Two-dimensional (2D) van der Waals (vdW) materials, sparked by graphene’s discovery \cite{Geim2007graphene}, have transformed materials science over the past two decades and
rapidly expanded, uncovering new prospects for both fundamental science and technological applications~\cite{Gupta2015recent,Wang2012Electronics,Geim2013heterostructures,Novoselov20162D,Ajayan20162D,Choi2017Recent,Manzeli20172D,Kolobov2016Two}. These atomically thin layers, such as 
transition metal dichalcogenides (TMDCs), hexagonal boron nitride (hBN), black phosphorus (BP), 
and transition metal monochalcogenides (TMMCs) have garnered significant attention due to their unique electronic, optical, and mechanical properties~\cite{Mak2010Atomically,Splendiani2010Emerging,Thygesen2012phonon,Rasmussen2015Computational,Haastrup2018Computational,Thygesen2015Dielectric,Massicotte20182D,Gupta2025TwoDimensional}. 
These materials exhibit some interesting phenomena such as bandgap tunability, 
spin-valve effects, proximity-induced effects, surface and interface effects, optoelectronic properties, and superconductivity~\cite{Liu2016vdW,Balents2020Superconductivity,Andrei2021moire,Gomez2022heterostructures,Stern2021Interfacial,Zhang2014Quantum}, making them ideal candidates for next-generation electronic and optoelectronic devices.

The remarkable properties of 2D vdW materials have spurred extensive theoretical investigations to understand and 
predict their behavior. Kohn-Sham (KS) density functional theory (DFT) has emerged as the state-of-the-art 
approach to study their fundamental characteristics~\cite{Kang2013band,Gong2013band,Zhu2011Giant,Liu2013tb,Kim2021thickness}. 
However, the accuracy of DFT calculations strongly depends on the choice of the exchange and correlation 
(XC)~\cite{Tran2021bandgap,patra2021efficient,Patra2020Electronic,Borlido2020XC,Tran2017Importance,patra2019efficient,GhoshJanaRauch2022,Excqual1,Exqual2}. While semilocal XC approximations offer reliable predictions for ground-state properties~\cite{Jana2018Assessing,patra2019relevance,sun2015strongly,jana2021szs,jana2019improving,patra2020way,jana2021accurate,Lebeda2024Balancing}, they often fail to capture excited-state phenomena such as bandgaps and optical absorption spectra~\cite{Neupane2021opening,Lebeda2023Right}. To address these limitations, it is necessary to go beyond semilocal approximations and incorporate Fock non-local 
exchange~\cite{Jana2020Improved,jana2022solid,janameta2018,jana2019screened,jana2018efficient,jana2020screened}. 
Within KS-DFT, this is achieved through hybrid density functionals, which provide reasonably accurate results for 
extended solids. While hybrid functionals improve accuracy for bulk materials, their fixed mixing parameters often 
struggle to describe the altered screening in lower-dimensional 2D vdW materials aterials~\cite{Wang2016Hybrid,Tran2021bandgap}, necessitating more tailored approaches. 

To overcome these challenges, many-body perturbation theory (MBPT)~\cite{Hedin1965GW,Hedin1969book,Onida2002electronic,hossain2022self} 
has emerged as a powerful alternative for accurately predicting the quasi-particle band structure and optical 
properties of 2D vdW materials. The Green’s function-based $GW$ 
approximation~\cite{Gao2022Numerical,hossain2021hybrid,Matthias2018Electronic,Espejo2013Band,Diana2013Optical,Yang2007Quasiparticle,Blase1995Quasiparticle,hossain2020transferability,Diana2016Screening,Cheiwchanchamnangij2012Quasiparticle,Golze2019GW,hossain2022self} 
and the Bethe-Salpeter equation 
(BSE)~\cite{Strinati1980Dynamical,Strinati1982Dynamical,Ramasubramaniam2019Transferable,Mishra2018Exciton,Camarasa2023Transferable,Marsili2021Spinorial,de2023optical,Lau2019Electronic,Attaccalite2011Real} 
offer more accurate descriptions of excited-state phenomena by incorporating dynamic screening and 
electron-hole interactions. While $GW$ captures the essential dynamic screening effects necessary for accurate 
bandgap calculations, its high computational cost remains a significant challenge. 
Among the different $GW$ variants, the non-self-consistent single-shot $G_0W_0$ method is often preferred for 
its balance between accuracy and computational efficiency. However, despite this advantage, $G_0W_0$ remains 
sensitive to the choice of the initial DFT starting  
point~\cite{Golze2019GW,Bruneval2013Benchmarking,Gant2022optimally,Jana2025Nonempirical}. To mitigate this limitation, 
dielectric-dependent hybrid functionals (DDHs) have gained attention as computationally efficient alternatives for 
predicting band gaps in 2D vdW solids. By replacing the fixed Fock mixing parameter with a material-specific 
dielectric 
constant~\cite{Ramasubramaniam2019Transferable,Camarasa2023Transferable,zhan2023nonempirical,jana2023simple,Ghosh2024accurate,Rani2024Thermoelectric}, DDHs bridge the gap between hybrid functionals and MBPT methods. However, determining 
the optimal mixing and range-separation parameters remains a key challenge, particularly for 2D systems, where 
conventional approaches used for 3D 
material~\cite{WeiGiaRigPas2018,wing2019comparing,jana2023simple,Gant2022optimally,OhadWingGant2022,WingHabeJonah2019,Ghosh2024accurate,Rani2024Thermoelectric,Jana2025Nonempirical,Jana2025metagga}, may not be directly 
applicable~\cite{Ramasubramaniam2019Transferable,Camarasa2023Transferable,Camarasa2023Transferable,zhan2023nonempirical}.

In this study, we employ screened-exchange dielectric-dependent range-separated hybrid (SE-DD-RSH) 
functional to systematic investigation of the bandgaps and optical properties of 2D  Van der Waals (vdW) materials. 
The proposed screened DDH functional achieves accuracy comparable to the highly demanding and computationally 
intensive $G_0W_0$ and BSE@$G_0W_0$ methods, offering a more efficient alternative for predicting electronic 
and optical properties. 

The paper is structured as follows: Section~\ref{theory} outlines the theoretical background of both existing and newly developed screened-exchange dielectric-dependent range-separated hybrid functionals. Section~\ref{results} presents a benchmark study and explores applications of the screened DDH functional for several prototypical vdW materials. Finally, Section~\ref{conclusion} summarizes our findings and key conclusions.

\section{Screened range-separated hybrid for layers}
\label{theory}
\subsection{Overview of screened range-separated hybrid}

The range-separated hybrid (RSH) density functional formalism introduces a refined treatment of exchange interactions 
by employing the Coulomb-attenuated method (CAM) to partition the two-electron 
operator~\cite{GeroBottCaraOnid2015,GeroBottCara2015,MicelChenIgor2018,ZhengGovoniGalli2019,GeroBottValeOnid2017,WeiGiaRigPas2018,HinuKumaTana2017,BrawGovoVoro2017,LiuCesaMart2020,WingHabeJonah2019,wing2019comparing,kronik2018dielectric,Ashwin2019Transferable,jana2023simple,Camarasa2023Transferable,Ghosh2024accurate,Jana2025metagga},
\begin{equation}
\frac{1}{r_{12}} = \frac{\alpha + (\gamma - \alpha) \operatorname{erf}(\mu r_{12})}{r_{12}} 
+ \frac{1 - \left[\alpha + (\gamma - \alpha) \operatorname{erf}(\mu r_{12})\right]}{r_{12}}.
\label{hy-eq1}
\end{equation}
where $\frac{1}{r_{12}}=\frac{1}{|{\bf{r}}_1-{\bf{r}}_2|}$ represents the Coulomb interaction between two electrons, 
$\alpha$ and $\gamma$, and $\mu$ are key parameters that define the separation between short- and long-range 
exchange interactions. By incorporating the Perdew-Burke-Ernzerhof (PBE)~\cite{perdewPRL96} XC and Fock exchange 
into Eq.~\ref{hy-eq1}, the exchange-correlation energy can be expressed as a sum of short-range ($sr$) and 
long-range ($lr$) components:

\begin{equation}
\begin{split}
E_{xc}^{RSH}(\alpha,\gamma;\mu) = (1-\alpha) E_x^{\text{PBE-sr},\mu} + \alpha E_x^{\text{Fock-sr},\mu} \\
+ (1-\gamma) E_x^{\text{PBE-lr},\mu} + \gamma E_x^{\text{Fock-lr},\mu} + E_c^{\text{PBE}}.
\end{split}
\label{hy-eq2}
\end{equation}
and the corresponding exchange-correlation potential is given by 
\begin{eqnarray}
V_{xc}^{RSH}({\bf{r}},{\bf{r}}',\alpha,\gamma;\mu)
&=& \left[\alpha+(\alpha-\gamma) \operatorname{erf}(\mu r)\right] V_x^{\text{Fock}}({\bf{r}},{\bf{r}}') \nonumber\\
&-& (\alpha-\gamma) V_x^{\text{PBE-sr},\mu}({\bf{r}}) \nonumber\\
&+& (1-\gamma) V_x^{\text{PBE}}({\bf{r}}) + V_c^{\text{PBE}}({\bf{r}}).
\label{hy-eq3}
\end{eqnarray}

Among the widely used short-range screened hybrids, the Heyd-Scuseria-Ernzerhof 
(HSE06)~\cite{heyd2003hybrid,heyd2004efficient} functional is particularly notable, typically adopting 
fixed parameters $\alpha=0.25$, $\gamma=0$, and $\mu=0.11$ Bohr$^{-1}$. However, alternative choices 
of $\alpha$ and $\mu$ have been explored in modified HSE like 
functionals~\cite{heyd2003hybrid,jana2020screened,jana2018efficient,jana2018meta,jana2019screened,Jana2020Improved}. 
A crucial refinement for accurately capturing excitonic effects in bulk solids is the proper asymptotic treatment of the Coulomb tail~\cite{tal2020accurate}. This can be achieved by setting $\gamma$ equal to the inverse of the 
high-frequency macroscopic static dielectric constant~\cite{WeiGiaRigPas2018,jana2023simple,Ghosh2024accurate}, 
where $\epsilon_{\infty}$ is the high-frequency macroscopic static dielectric function or ion-clamped static 
(optical) dielectric function. This dielectric-dependent hybrid (DDH) approach improves the description of 
electronic screening, but its transferability across different material systems remains an open 
challenge~\cite{Ashwin2019Transferable}. In practice, additional strategies may be necessary to determine 
optimal values of $\alpha$, $\gamma$, and $\mu$. 

As evident from Eqs.~\ref{hy-eq2} and ~\ref{hy-eq3}, beyond the choice of $\alpha$, accurate determination of 
$\epsilon_{\infty}$ and $\mu$ is critical for achieving predictive RSH calculations.  Several 
methodologies exist to determine these parameters, which will be discussed in the following sections. 

\subsection{Strategies for determining the parameters and Present Consideration}
First, we consider the bulk systems (for example, simple cubic bulk). A finite electronic field is applied to determine the dielectric tensor. Because the bulk system is homogeneous, it is straightforward to obtain the macroscopic dielectric constant as,
\begin{equation} \epsilon^{\text{3D}}_\infty=\frac{1}{3}\sum_{i=1}^{3}\epsilon^{\text{3D}}_{\infty,i}~,
\label{eq4}
\end{equation}
where $\epsilon^{\text{3D}}_{\infty,i}~~(i=1,2,3)$ are the diagonal elements of the dielectric tensor. However, in layered materials, the dielectric response is anisotropic due to the coexistence of distinct materials within the system. Consequently, the assumption of homogeneity used for bulk solids is no longer valid. Because of this anisotropy of van der Waals (vdW) layer materials, the in-plane ($\parallel$) and out-of-plane ($\perp$) components of the dielectric tensor exhibit significant differences. In such cases, a more sophisticated approach is required to determine the dielectric response accurately.

In the following, we propose the construction of the parameters of the screened-DDH for two-dimensional vdW systems. Accurate calculation of the dielectric response requires the elimination of the vacuum contribution. For layered materials, the macroscopic static dielectric tensor is first calculated using a supercell approach with a sufﬁciently large vacuum layer to avoid the non-physical interaction between periodic repetitions of layers. The macroscopic dielectric tensor of the supercell (SC), $\epsilon^{SC}_\infty$ is determined using the reciprocal space expression~\cite{Gajdo2006linear,Nunes2001berry,Souza2002first},
\begin{equation}
\epsilon^{\text{SC}}_{{\bf{G}},{\bf{G'}}}({\bf{q}},\omega)=\delta_{{\bf{G}},{\bf{G'}}}-v_c\chi^{\text{SC}}_{{\bf{G}},{\bf{G'}}}({\bf{q}},\omega)~.
\label{eq-epsilon-rpa}
\end{equation}
where $v_c=\frac{4\pi}{|{\bf{G+q}}||{\bf{G'+q}}|}$ is the Hartree potential and
\begin{equation}
 \chi^{\text{SC}}_{{\bf{G}},{\bf{G'}}}=\chi^{0,\text{SC}}_{{\bf{G}},{\bf{G'}}}+\chi^{0,SC}_{{\bf{G}},{\bf{G'}}}\Big(v_c+f_{xc}({\bf{q}},\omega)\Big)\chi^{\text{SC}}_{{\bf{G}},{\bf{G'}}}~,
\end{equation}
with $f_{xc}({\bf{q}},\omega)$ being the
XC kernel, $\chi^{SC}_{{\bf{G}},{\bf{G'}}}({\bf{q}},\omega)$ is the reducible polarizability, and $\chi^0_{{\bf{G}},{\bf{G'}}}({\bf{q}},\omega)$ being the irreducible polarizability matrix of the supercell. However, often the effect of $f_{xc}({\bf{q}},\omega)$ is neglected and Eq.~\ref{eq-epsilon-rpa} is then referred as RPA. Our interest lies in the inverse of the macroscopic dielectric matrix $[\epsilon^{SC}_\infty({\bf{q}},\omega)]^{-1}$, which is obtained from the head of the inversion of the full microscopic dielectric tensor as~\cite{Gajdo2006linear},
\begin{equation}
[\epsilon^{\text{SC}}_\infty({\bf{q}},\omega)]^{-1}=\lim_{{\bf{q}}\to 0}[\epsilon^{\text{SC}}_{0,0}({\bf{q}},\omega)]^{-1} \; .
\end{equation}

Typically, for layer 2D materials, the static microscopic dielectric tensor, $\epsilon^{\text{SC}}_{\infty}$ has following matrix form,
\begin{equation}
\overleftrightarrow{\epsilon^{\text{SC}}_\infty} =
\begin{pmatrix}
\epsilon^{\text{SC}}_{\infty\parallel} & 0 & 0 \\
0 & \epsilon^{\text{SC}}_{\infty\parallel} & 0 \\
0 & 0 & \epsilon^{\text{SC}}_{\infty\perp}
\end{pmatrix}.
\end{equation}
where $\epsilon_{\infty,\parallel}^{SC}$ is the in-plane dielectric constant and $\epsilon_{\infty,\perp}^{\text{SC}}$ is the out-of-plane dielectric constant of the supercell, which includes contributions from both the 2D material and
the vacuum. Typically,  the density functional perturbation theory (DFPT)~\cite{Baroni2001phonons} is used to calculate the
susceptibility tensor of the supercell. We use RPA@PBE to determine the dielectric tensor where the local field effect is included within the Hartree approximation~\cite{Gajdo2006linear}.

We adopt the following strategy to determine the dielectric constants for two-dimensional layer materials, including vacuum: \\

(i) We
ﬁrst calculate the supercell macroscopic dielectric constants $\epsilon_\infty^{SC}$ for both the in-plane $\epsilon_{\infty,\parallel}^{SC}$ and out-of-plane $\epsilon_{\infty,\perp}^{SC}$ using RPA@PBE which includes contributions from both the 2D material and
the vacuum.\\

\begin{figure*}
    \centering
    \begin{subfigure}
        \centering
        \includegraphics[width=8.1 cm]{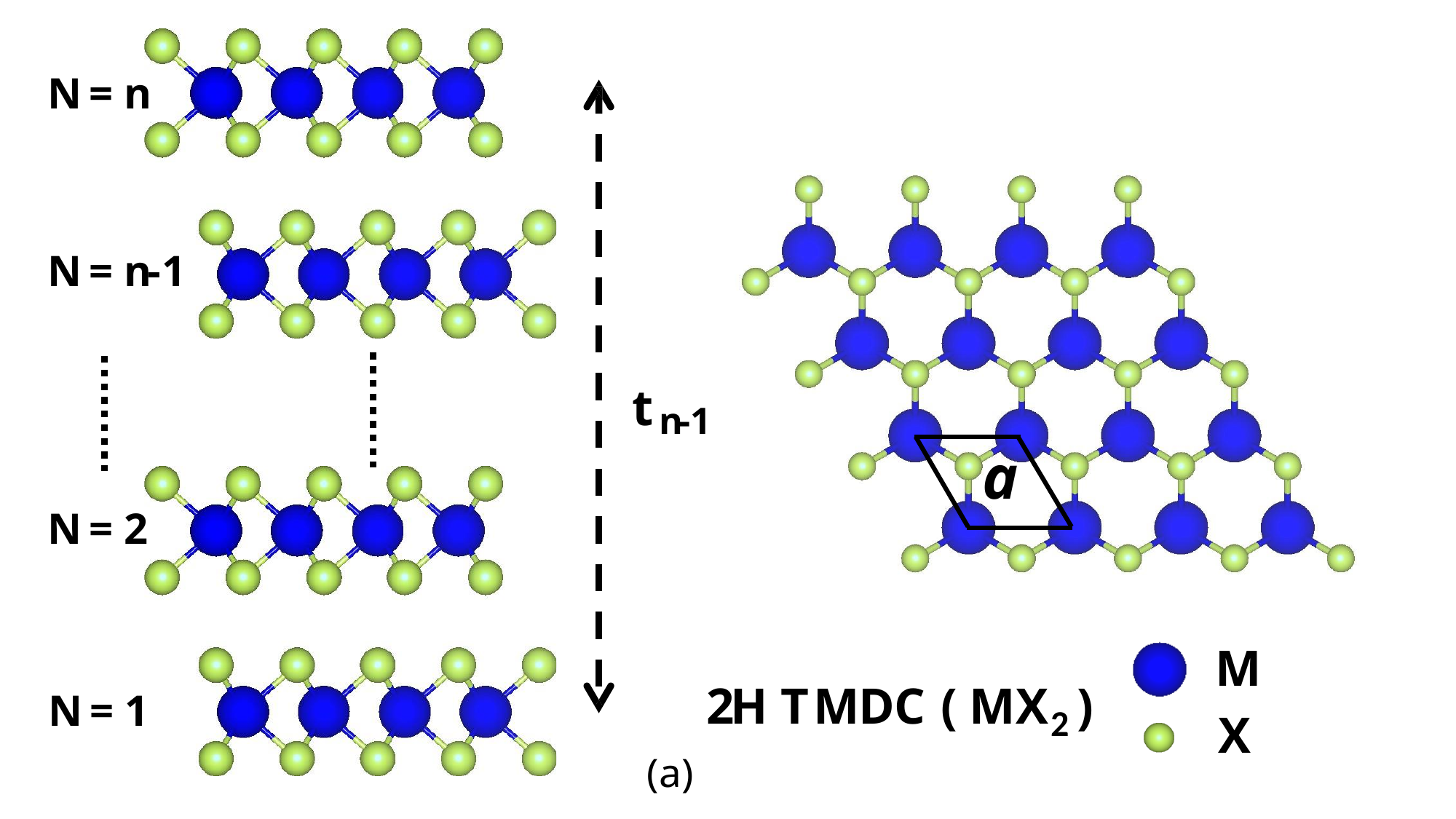}
                       \includegraphics[width=8.1 cm]{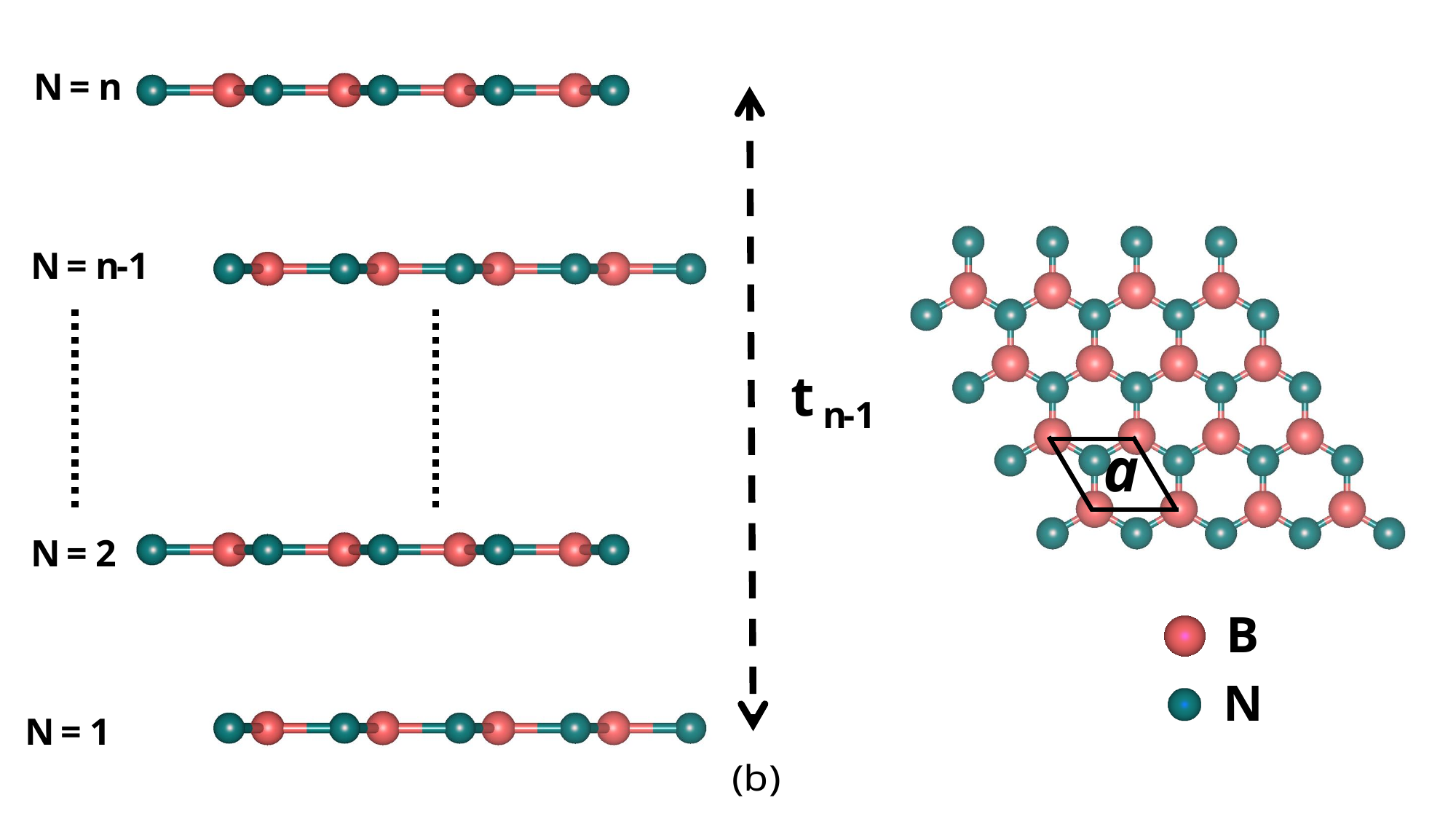}
    \end{subfigure}
    \begin{subfigure}
        \centering
        \includegraphics[width=8.1 cm]{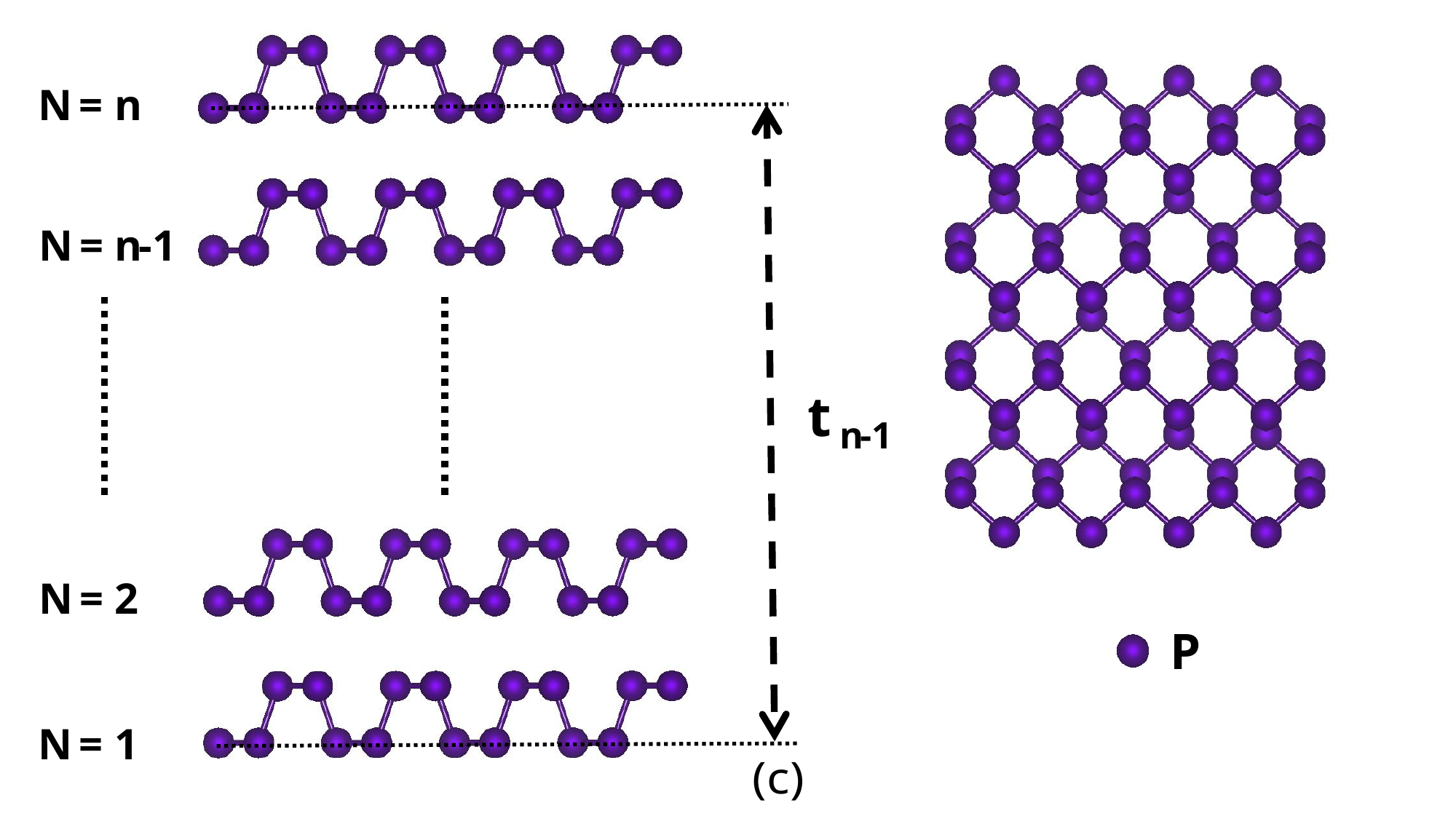}
        \includegraphics[width=8.1 cm]
        {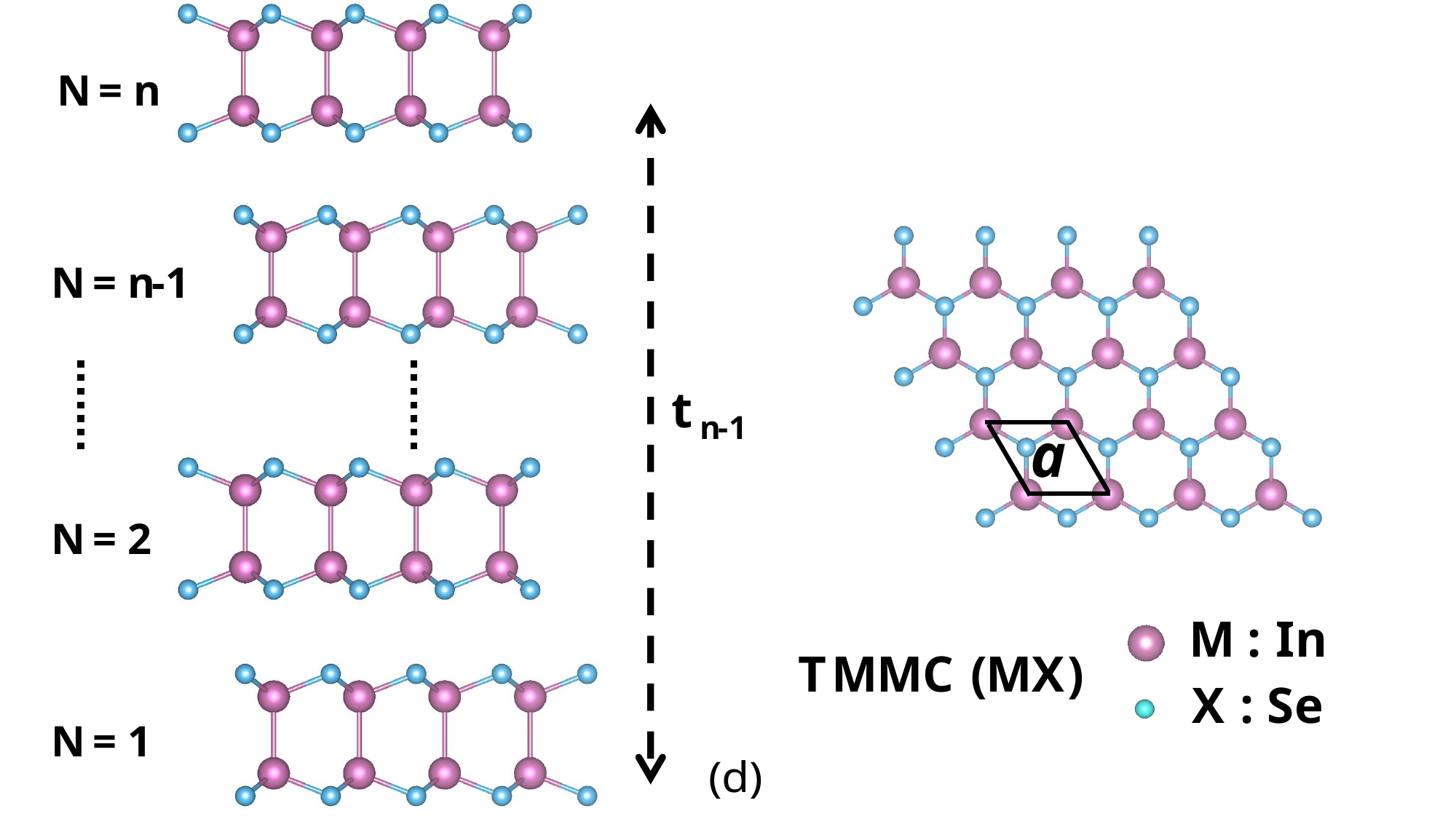}
    \end{subfigure}
    \caption{Multilayer $AB$ staking configuration of 2H TMDC of MX$_2$ (a), $AA'$ stacking of $h$BN (b), $AB$ stacking of BP (c), and $AB$ stacking of $\beta-$InSe (d). Here, $a$ is the in-plane lattice constant, and $t_{n-1}$ is the thickness (or the distance) between the metal atoms of $n+1$ layers when $n$ layers are considered.}
    \label{structure-fig}
\end{figure*}

(ii) The in-plane and out-of-plane dielectric constant for 2D materials is then obtained by rescaling the supercell dielectric constant as,
\begin{eqnarray}
\epsilon_{\infty\parallel}^{\text{2D}} &=& 1 + \frac{c}{t} (\epsilon_{\infty\parallel}^{\text{SC}} - 1) \nonumber\\
\epsilon_{\infty\perp}^{\text{2D}} &=& \left[ 1 + \frac{c}{t} \left( \frac{1}{\epsilon_{\infty\perp}^{\text{SC}}} - 1 \right) \right]^{-1}.
\label{eq-rescale}
\end{eqnarray}
%
Here, $c$ is the supercell height or length perpendicular to the 2D material layers, and $t$ is the thickness of the 2D material layers (see Fig.~\ref{structure-fig} for details). The relation of Eq.~\ref{eq-rescale} is obtained from {\it{principle of
equivalent capacitance}}~\cite{Aspnes1982local,Laturia2018dielectric} (See Appendix~\ref{Appendix-A} for details.) 
The important fact of this consideration is that the rescaled $\epsilon^{2D}_{\infty\parallel}$ and $\epsilon^{2D}_{\infty\perp}$ remain almost independent with the different supercell height, $c$. In all our calculations we consider $\epsilon_{vacuum} = 1$.  

Note also for a free-standing monolayer or strict 2D limits, i.e., for large (infinite) separation between layers, that $\epsilon^{SC}_{\infty\parallel}\to 1$, $\epsilon^{SC}_{\infty\perp}\to 1$, resulting $\epsilon^{2D}_{\infty\parallel}\to 1$, $\epsilon^{2D}_{\infty\perp}\to 1$~\cite{Laturia2018dielectric}. But Eq.~\ref{eq-rescale} is more valid and physical for finite thickness, which our present work uses to construct the DDH for 2D layer materials.


(iii) Finally, to solve the DDH equation, one needs to construct the effective dielectric constant, $\epsilon^{2D}_{\infty}$. The following choices can be adopted:\\

\begin{itemize}
    \item For the bulk system, $c=t$, which corresponds to $\epsilon^{2D}_{\infty\parallel}=\epsilon^{SC}_{\infty\parallel}$, and $\epsilon^{2D}_{\infty\perp}=\epsilon^{SC}_{\infty\perp}$, implies the effective dielectric constants becomes the averaging over diagonal elements of the static dielectric tensor,
\begin{equation}
[\epsilon_{\infty}^{\text{2D, eff}}]_{\text{bulk}} = \frac{2\epsilon_{\infty\parallel}^{\text{2D}} + \epsilon_{\infty\perp}^{\text{2D}}}{3}.
\label{die-eff1}
\end{equation}
where we assume $[\epsilon_{\infty\parallel}^{\text{2D}}]_{xx} = [\epsilon_{\infty\parallel}^{\text{2D}}]_{yy}
$. Thus Eq.~\ref{die-eff1} is, in effect, the same as Eq.~\ref{eq4}.
\end{itemize}

\begin{itemize}
    \item However, for layers, such as monolayer or bilayer one has to consider different strategies for effective $\epsilon^{2D, eff}_{\infty}$. Following Appendix~\ref{Appendix-A}, for a monolayer or bilayer of the 2D vdW material, we consider the effective dielectric constant as,
\begin{equation}
[\epsilon_{\infty}^{\text{2D, eff}}]_{\text{layers}} = \sqrt{\epsilon_{\infty\parallel}^{\text{2D}} \epsilon_{\infty\perp}^{\text{2D}}}.
\label{die-eff2}
\end{equation}

Eventually, Eq.~\ref{die-eff2} is also used in refs.~\cite{VanTuan2018coulomb,Cheiwchanchamnangij2012Quasiparticle} as the effective dielectric constant for Coulomb interaction in the plane for monolayer-TMDs.
\end{itemize}

We use Eq.~\ref{die-eff1} (in case of bulk) and Eq.~\ref{die-eff2} (in case of a monolayer) in our screened DDH potential (or exchange energy functional) to calculate the 2D vdW material properties. Finally, using Eq.~\ref{hy-eq3}, the generalized potential energy of the screened-exchange dielectric-dependent range-separated hybrid (SE-DD-RSH) becomes,
\begin{eqnarray}
V_{xc}^{\text{SE-DD-RSH}}({\bf{r}},{\bf{r'}};\mu) 
&=& \left[1 - \left(1 - \left[\epsilon_{\infty}^{\text{2D, eff}}\right]^{-1} \right) \operatorname{erf}(\mu r) \right] V_x^{\text{Fock}}({\bf{r}},{\bf{r'}}) \nonumber\\
&-& \left(1 - \left[\epsilon_{\infty}^{\text{2D, eff}}\right]^{-1} \right) V_x^{\text{PBE-sl},\mu}({\bf{r}}) \nonumber\\
&+& \left(1 - \left[\epsilon_{\infty}^{\text{2D, eff}}\right]^{-1} \right) V_x^{\text{PBE}}({\bf{r}}) + V_c^{\text{PBE}}({\bf{r}})~,
\label{eq-ddh-eff}
\end{eqnarray}
where we choose $\alpha=1$ in Eq.~\ref{hy-eq3}. Here, the model dielectric function,
\begin{eqnarray}
 \epsilon_m=1 - \left(1 - \left[\epsilon_{\infty}^{\text{2D, eff}}\right]^{-1} \right) \operatorname{erf}(\mu r)~.   
\end{eqnarray}
For free-standing monolayer or strict 2D limit $\epsilon_{\infty}^{\text{2D, eff}}\to [\epsilon_{\infty}^{\text{2D, eff}}]_{layers}\to 1$, implies $\epsilon_m\to 1$. This also implies  
free-standing monolayer or strict 2D limit the long-range exchange potential $\sim \frac{1}{q^2}$.

Regarding $\mu$, we consider $\mu=\mu_{bulk}=\mu_{eff}^{fit}$ from ref.~\cite{jana2023simple,Ghosh2024accurate} for bulk, given by 
\begin{equation}
 \mu_{bulk}=\mu_{eff}^{fit}= \frac{a_1}{\langle r_s \rangle} + \frac{a_2 \langle r_s \rangle}{1 + a_3 \langle r_s \rangle^2}~,
 \label{eq-theo-secb-19}
\end{equation}
with $a_1= 1.91718$, $a_2= -0.02817$, $a_3=0.14954$, and
\begin{equation}
 \langle r_s \rangle=\frac{1}{V_{cell}}\int_{cell} \Big(\frac{3}{4\pi(n_{\uparrow}({\bf{r'}})+n_{\downarrow}({\bf{r'}}))}\Big)^{1/3}~d^3r'~.
 \label{eq-theo-secb-13}
\end{equation}

However, for the monolayer, the strategies of ref.~\cite{jana2023simple,Ghosh2024accurate} are not convenient because of the involvement of the volume of the unit cell. In contrast, we consider the following function form to be useful,
\begin{equation}
\mu_{\text{layers}} = f\left( \left[ \epsilon_{\infty}^{\text{2D, eff}} \right]_{\text{layers}}^{-1} \right).
\end{equation}
and out of the  different forms proposed for bulk solids (see Table 1 in the refs.~\cite{LiuCesaMart2020}), we consider the following generalized form with a fitting parameter for the 2D monolayer of the vdW systems,  
\begin{equation}
\mu_{\text{layers}} = a N_e^{1/3} \left[ 1 - \left( \epsilon_{\infty}^{\text{2D, eff}} \right)_{\text{layers}}^{-1} \right]^{-1/2}.
\label{2d-mu}
\end{equation}
where $N_e$ represents the number of the valence electrons. For instance, in the case of MoX$_2$ or WX$_2$
($X=S/Se/Te$) $N_e=N_e^{Mo/W}=6$; for InSe, $N_e=N_e^{In}=3$, and for hexagonal boron nitride ($h$BN), $N_e=N_e^{B}=3$. The parameter, $a=0.3$ is obtained from fitting with the reference $G_0W_0$ band gap data and provides a well-balanced estimation of the band gap across various monolayer materials. As stated before, we also note that Eq.~\ref{2d-mu} is also used previously in the case of bulk solids~\cite{LiuCesaMart2020}, which is modified or scaled by $a$ in the present case for application to the lower-dimensional materials. Once the parameter $a$ is determined and fixed, no additional fitting is required for other systems outside the test set. Since $\mu_{\text{layers}}$ depends solely on the dielectric constant of 2D systems (assuming 
$a$ remains fixed), it can be readily evaluated without further fitting.

\subsection{Other recent strategies}

We also recall some recent progress in constructing the screened-exchange functionals for 2D materials, such as,

(a) For heterogeneous systems such as interfaces and surfaces, Zhan et al.~\cite{zhan2023nonempirical} proposed the same form as Eq.~\ref{eq-ddh-eff}, but a local, spatially-dependent mixing fraction, $\gamma({\bf{r}},{\bf{r}}')$ is used,
\begin{equation}
\gamma({\bf{r}},{\bf{r}}')=\frac{1}{\sqrt{\epsilon_\infty({\bf{r}})\epsilon_\infty({\bf{r}'})}}=\frac{1}{\sqrt{\epsilon_\infty^{\text{2D}}({\bf{r}},{\bf{r}}')}}~,
\label{galli-1}
\end{equation}
with $\epsilon_\infty({\bf{r}})$ being the local dielectric function. Their screened-exchange range-separated hybrid (SE-RSH) also takes the form $\alpha=1$, and $\mu=\mu({\bf{r}})$ is also a position-dependent function, and it is obtained in a self-consistent manner~\cite{zhan2023nonempirical}. The screened-dielectric function proposed by Zhan et al. ~\cite{zhan2023nonempirical} is fully nonempirical, and all the parameters are position-dependent. Notably, the SE-RSH function is the most generalized form, suitable for complex materials such as bulk, heterogeneous materials, interfaces, and two-dimensional layered solids. The cost has to be paid in terms of an extra effort to implement the more complex local dielectric function, $\epsilon({\bf{r}})$, and the local screening function $\mu({\bf{r}})$. We recall that the local dielectric function is determined in real space using the Wannier basis. 

(b) An optimal tuning of parameters related to the two-dimensional vdW systems is proposed in various papers published by Ramasubramaniam et al. ~\cite{Ashwin2019Transferable,Camarasa2023Transferable}. This is based on two main facts: (i) Because $\epsilon_\infty(q)$ tends to $\epsilon_\infty(q)=1+{\mathcal{O}}(q)$ for a strict 2D limit~\cite{Huser2013dielectric}, using Eq.~\ref{hy-eq3}, the potential energy functional of their screened-range-separated hybrid (SRSH) looks like
\begin{eqnarray}
 V_{xc}^{SRSH}({\bf{r}}, {\bf{r}}',\alpha;\mu)
&=& \alpha [1+\operatorname{erf}(\mu r)] V_x^{\text{Fock}}({\bf{r}}, {\bf{r}}') \nonumber\\
&+& (1-\alpha) V_x^{\text{PBE-sr},\mu}({\bf{r}}) + V_c^{\text{PBE}}({\bf{r}}).
\label{hy-kronik}
\end{eqnarray}
(ii) In Eq.~\ref{hy-kronik}, then a scanning over the $\alpha-\mu$ parameters is performed such that the corresponding SRSH band gaps match with the reference quasi-particle $G_0W_0$ result by imposing the constraints that the difference between the quasiparticle band gaps and SRSH band gaps at a particular $K$ point becomes optimal, i.e.,
\begin{eqnarray}
\Delta E_g^{\alpha,\mu} = \arg\min_{\alpha,\mu} |E_g^{G_0W_0} - E_g^{SRSH}|.
\label{hy-kronik2}
\end{eqnarray}
As noted from ref.~\cite{Camarasa2023Transferable}, the corresponding pairs of $\alpha$ and $\mu$ are not unique and depend on the materials. Furthermore, the previous band gap calculations of the quasiparticle $GW$ are required. However, as also noted in ref.~\cite{Camarasa2023Transferable}, a unique or optimal set of parameters ($\alpha^*,\mu^*$) is obtained by simultaneously optimizing the bulk and monolayer band gap by imposing the constraint of Eq.~\ref{hy-kronik2}, which is ultimately used to compute the electronic and optical properties of 2D vdW materials. Other than this optimal tuning procedure, very recently, a non-empirical screened range-separated hybrid functional is also proposed using a Wannier-localized optimally-tuned procedure~\cite{Gomez2024Excitations}.

In the present scenario, our proposed method takes the same form as that of Zhan et al.~\cite{zhan2023nonempirical}, only instead of the non-local $\epsilon^{-1}({\bf{r}},{\bf{r}}')$, we have used $\epsilon^{2D, eff}_{\infty}$, a fixed effective value. Thus, the present method can be thought of as similar to the most generalized form of ref.~\cite{zhan2023nonempirical}. Moreover, the present approach differs from case (ii) in that, for bulk systems, our method is entirely non-empirical, whereas for 2D layered systems, only the dielectric constant $\epsilon$
is required, assuming $a=0.3$ is fixed to a specific value.

\section{Results and Discussions}
\label{results}

\begin{table*}
    \caption{\label{tab-band-gaps} Direct band gaps (eV) of bulk and monolayer vdW materials.
    For MoX$_2$, WX$_2$, and $h$BN, the direct band gaps are computed at the $K_v\to K_c$ transition;
    while for BP and InSe, at $\Gamma_v\to \Gamma_c$ is considered. Spin-orbit coupling (SOC) effects,
    derived from PBE, are subtracted across methods. Values in parentheses indicate deviations from
    corresponding $G_0W_0$ results.  
    }

    \centering
    \scriptsize 
    \setlength{\tabcolsep}{4pt} 

    \resizebox{\textwidth}{!}{ 
    \begin{tabular}{l c c c c c c c c c c}
        \toprule
        \textbf{Material} & \textbf{Transition} & \textbf{$k$-points} & \boldmath{$\gamma = [\epsilon_{\infty}]^{-1}$} & \boldmath{$\mu$ (\AA$^{-1}$)} & \boldmath{$E_g^{PBE}$} & \boldmath{$E_g^{HSE06}$} & \boldmath{$E_g^{SE-DD-RSH}$} & \boldmath{$E_g^{G_0W_0}$} & \boldmath{$\Delta$SOC$^b$} \\
        \midrule
        \multicolumn{10}{c}{\textbf{Bulk}} \\[0.1cm]
MoS$_2$	&	$K_v\to K_c$	&	$12\times 12\times 4$	&	0.0867	&	1.739	&	1.69 (-0.46)	&	2.14 (0.00)	&	2.10 (-0.04)	&	2.14	&	0.03	\\
MoSe$_2$	&	$K_v\to K_c$	&	$12\times 12\times 4$	&	0.0760	&	1.720	&	1.50 (-0.54)	&	1.94 (-0.10)	&	1.86 (0.02)	&	1.88	&	-0.03	\\
MoTe$_2$	&	$K_v\to K_c$	&	$12\times 12\times 4$	&	0.0612	&	1.720	&	1.13 (-0.51)	&	1.55 (-0.08)	&	1.42 (0.00)	&	1.42	&	-0.05	\\
WS$_2$	&	$K_v\to K_c$	&	$12\times 12\times 4$	&	0.0946	&	1.814	&	1.67 (-0.54)	&	2.15 (-0.06)	&	2.12 (-0.09)	&	2.21	&	0.19	\\
WSe$_2$	&	$K_v\to K_c$	&	$12\times 12\times 4$	&	0.0803	&	1.757	&	1.33 (-0.50)	&	1.82 (-0.01)	&	1.71 (-0.12)	&	1.83	&	0.22	\\
WTe$_2$	&	$K_v\to K_c$	&	$12\times 12\times 4$	&	0.0615	&	1.739	&	0.81 (-0.35)	&	1.22 (0.06)	&	1.10 (-0.06)	&	1.16	&	0.28	\\
$h$BN	&	$K_v\to K_c$	&	$12\times 12\times 4$	&	0.2561	&	1.537	&	4.96 (-1.68)	&	6.43 (-0.21)	&	7.59 (0.94)	&	6.64 (7.32$^a$) 	&	0.00	\\
BP	&	$\Gamma_v\to \Gamma_c$	&	$9\times 9\times 4$	&	0.1080	&	1.587	&	0.03 (-0.68)	&	0.52 (-0.18)	&	0.49 (-0.22)	&	0.71	&	0.00	\\
InSe	&	$\Gamma_v\to \Gamma_c$	&	$9\times 9\times 4$	&	0.1137	&	1.285	&	0.38 (-0.93)	&	1.10 (-0.21)	&	1.21 (-0.15)	&	1.32	&	0.03	\\
        \midrule
        \multicolumn{10}{c}{\textbf{Layers}} \\[0.1cm]
MoS$_2$	&	$K_v\to K_c$	&	$15\times 15\times 1$	&	0.1144	&	1.095	&	1.70	(-0.85)	&	2.16	(-0.39)	&	2.31	(-0.24)	&	2.55	&	0.08	&	1L	\\
	&		&		&	0.1076	&	1.091	&	1.70	(-0.66)	&	2.15	(-0.21)	&	2.28	(-0.08)	&	2.36	&	0.04	&	2L	\\
MoSe$_2$	&	$K_v\to K_c$	&	$15\times 15\times 1$	&	0.0996	&	1.086	&	1.48	(-0.75)	&	1.93	(-0.30)	&	2.03	(-0.20)	&	2.23	&	0.11	&	1L	\\
	&		&		&	0.0930	&	1.082	&	1.48	(-0.58)	&	1.93	(-0.13)	&	2.00	(-0.06)	&	2.06	&	0.06	&	2L	\\
MoTe$_2$	&	$K_v\to K_c$	&	$15\times 15\times 1$	&	0.0768	&	1.072	&	1.07	(-0.61)	&	1.51	(-0.17)	&	1.54	(-0.14)	&	1.68	&	0.13	&	1L	\\
	&		&		&	0.0707	&	1.069	&	1.08	(-0.45)	&	1.50	(-0.03)	&	1.51	(-0.02)	&	1.53	&	0.08	&	2L	\\
WS$_2$	&	$K_v\to K_c$	&	$15\times 15\times 1$	&	0.1206	&	1.099	&	1.72	(-1.01)	&	2.21	(-0.52)	&	2.42	(-0.32)	&	2.73	&	0.26	&	1L	\\
	&		&		&	0.1153	&	1.095	&	1.73	(-0.81)	&	2.21	(-0.33)	&	2.40	(-0.14)	&	2.54	&	0.21	&	2L	\\
WSe$_2$	&	$K_v\to K_c$	&	$15\times 15\times 1$	&	0.1052	&	1.089	&	1.42	(-0.88)	&	1.89	(-0.41)	&	2.02	(-0.28)	&	2.30	&	0.29	&	1L	\\
	&		&		&	0.0996	&	1.086	&	1.43	(-0.71)	&	1.89	(-0.25)	&	2.00	(-0.14)	&	2.14	&	0.24	&	2L	\\
WTe$_2$	&	$K_v\to K_c$	&	$15\times 15\times 1$	&	0.0800	&	1.074	&	0.90	(-0.70)	&	1.33	(-0.27)	&	1.37	(-0.23)	&	1.60	&	0.32	&	1L	\\
	&		&		&	0.0741	&	1.071	&	0.91	(-0.54)	&	1.32	(-0.13)	&	1.33	(-0.12)	&	1.45	&	0.25	&	2L	\\
$h$BN	&	$K_v\to K_c$	&	$15\times 15\times 1$	&	0.3043	&	0.980	&	4.70	(-2.72)	&	6.11	(-1.31)	&	8.35	(0.94)	&	7.42 (8.32$^a$)	&	0.00	&	1L	\\
	&		&		&	0.3001	&	0.977	&	4.74	(-2.36)	&	6.15	(-0.94)	&	8.38	(1.28)	&	7.10 (7.87$^a$)	&	0.00	&	2L	\\
BP	&	$K_v\to K_c$	&	$15\times 15\times 1$	&	0.1046	&	1.024	&	0.67 (-0.84)		&	1.32 (-0.19)		&	1.48 (-0.03)		&1.51		&	0.00	&	1L	\\
	&		&		&	0.0767	&	1.009	&	0.29 (-0.60)		&	0.88 (-0.01)		&	0.90 (0.01)		&0.89		&	0.00	&	2L	\\
InSe	&	$K_v\to K_c$	&	$15\times 15\times 1$	&	0.1507	&	0.887	&	1.71	(-1.32)	&	2.49	(-0.55)	&	3.22	(0.18)	&	3.04	&	0.05	&	1L	\\
	&		&		&	0.1416	&	0.882	&	1.16	(-1.09)	&	1.87	(-0.38)	&	2.50	(0.25)	&	2.25	&	0.02	&	2L	\\
        \bottomrule
    \end{tabular}
    }
    \footnotetext{partially self-consistent $GW$ ($scGW_0$)
    calculation.}
\footnotetext{$\Delta~SOC=E_g^{PBE}-E_g^{PBE-SOC}$}
\end{table*}

\begin{figure*}
    \centering
    \begin{subfigure}
        \centering
        \includegraphics[width=8 cm]{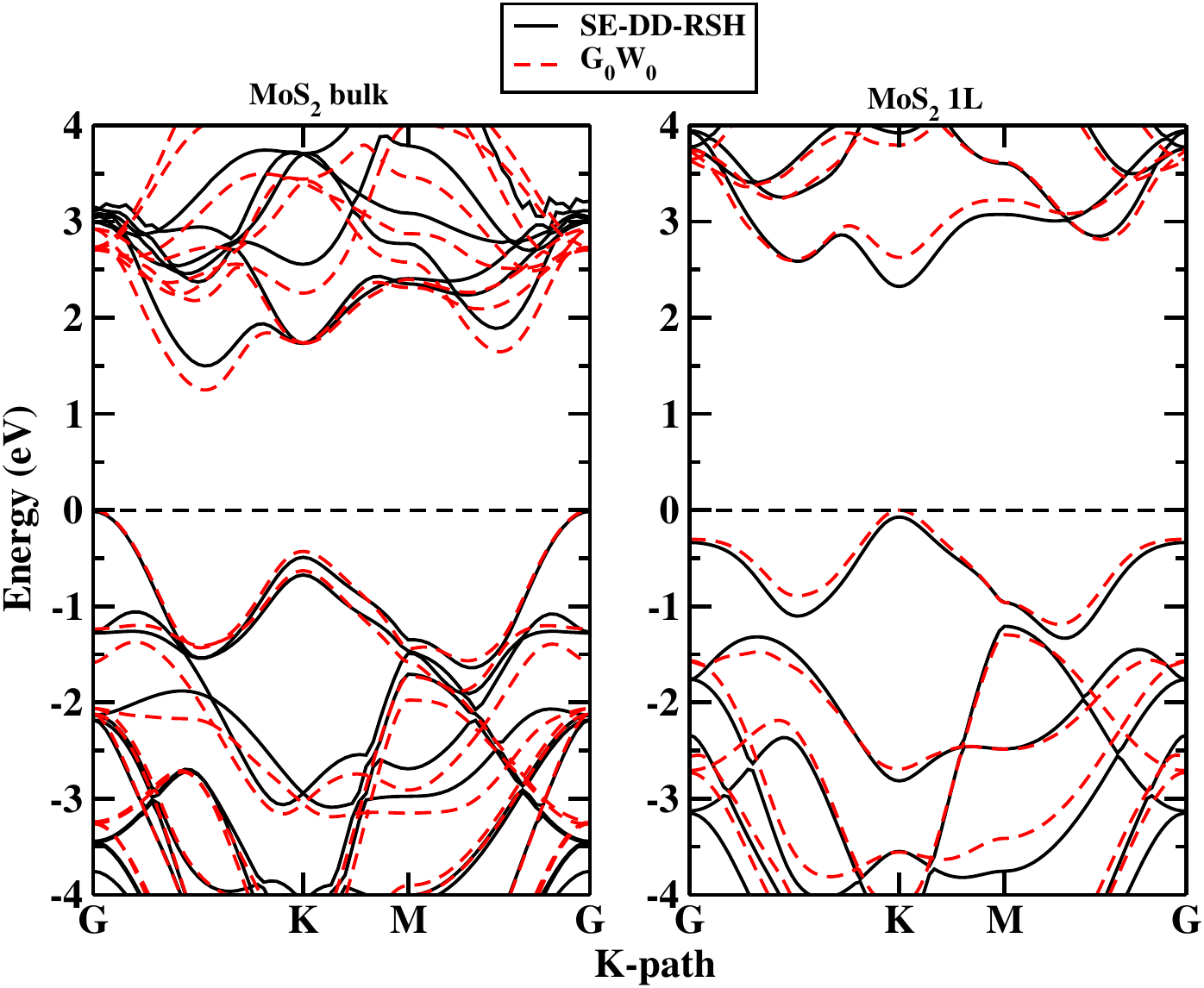}
                       \includegraphics[width=8 cm]{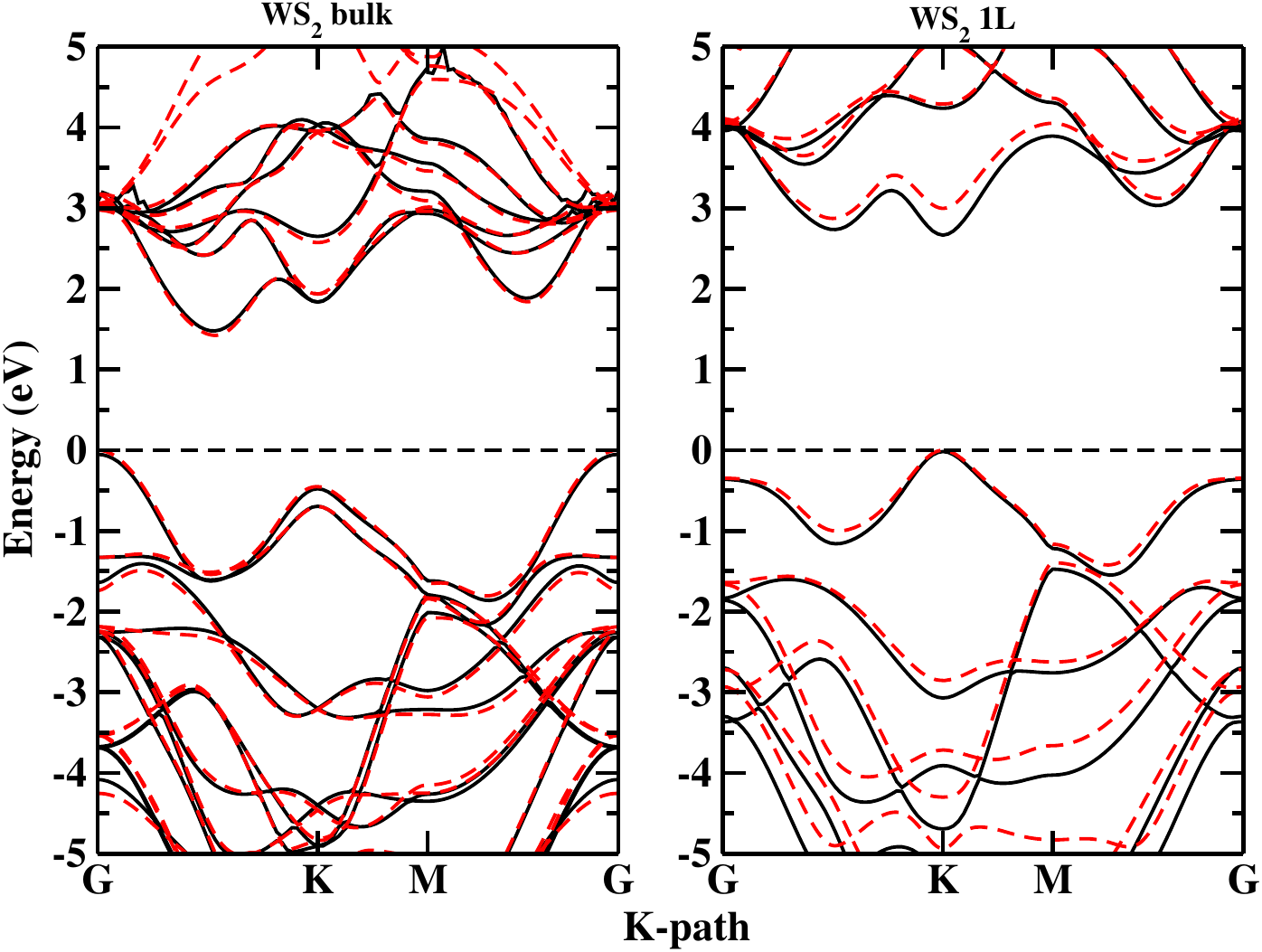}
    \end{subfigure}
    \begin{subfigure}
        \centering
        \includegraphics[width=8 cm]{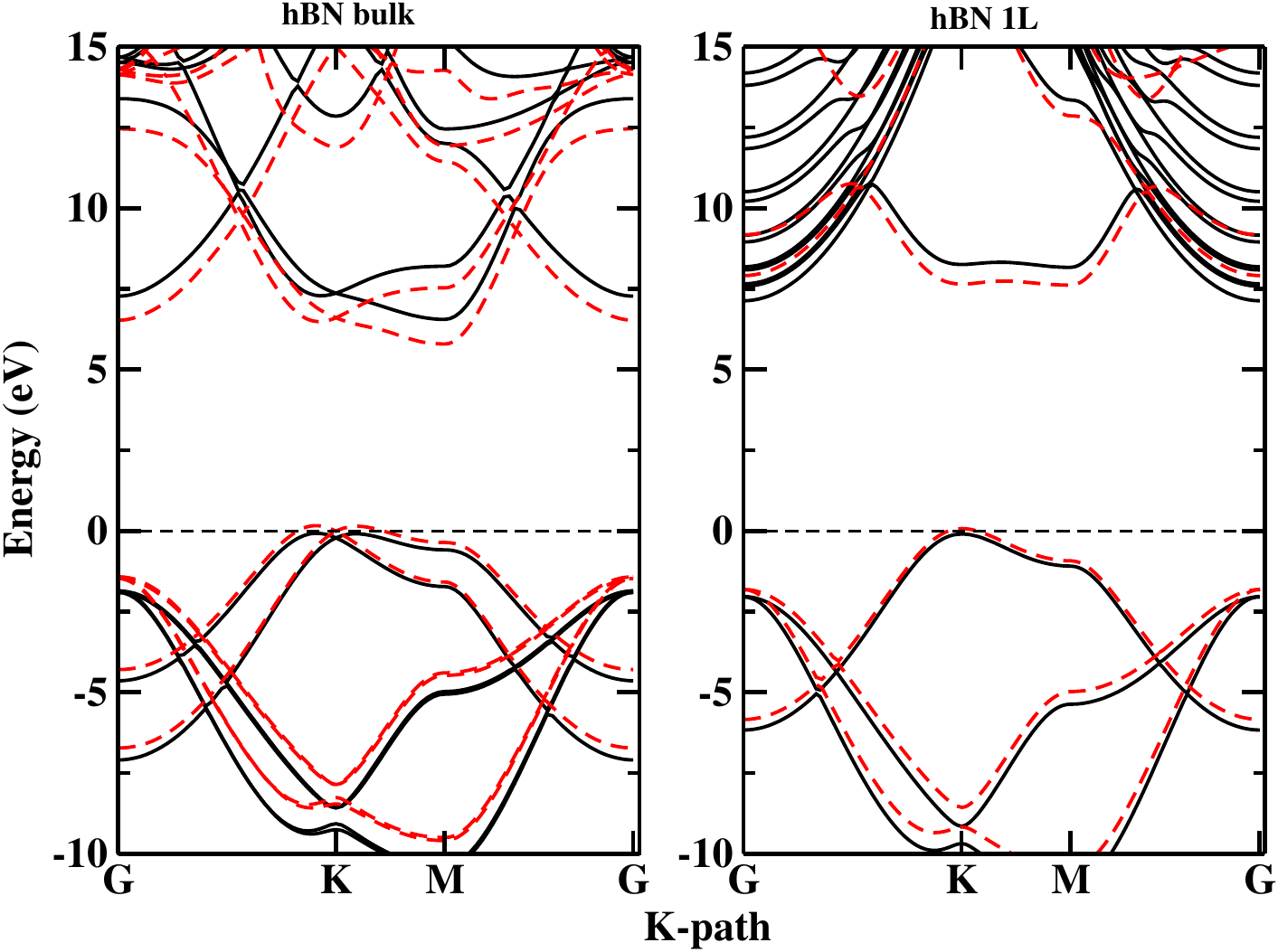}
        \includegraphics[width=8 cm]
        {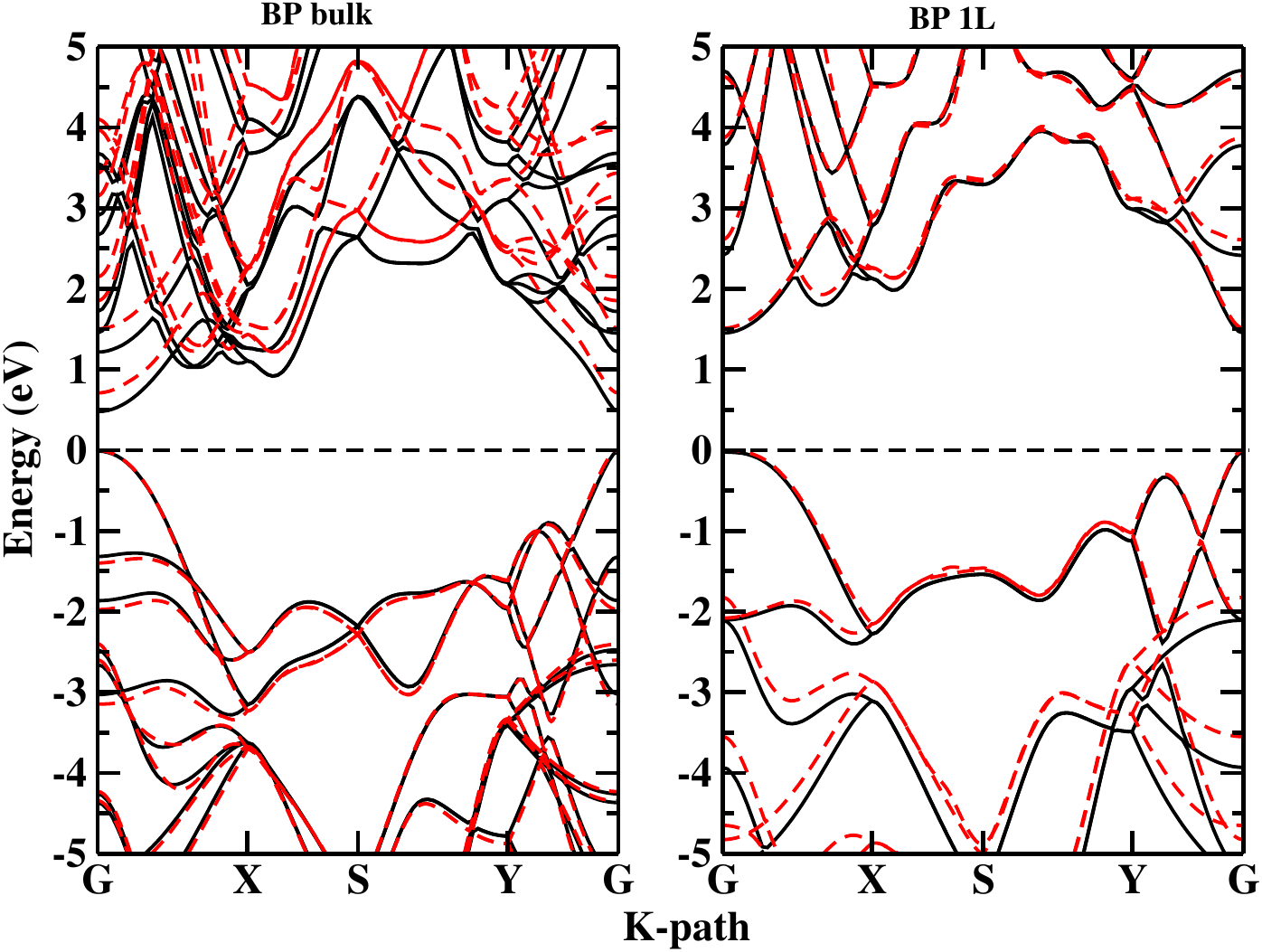}        
    \end{subfigure}
    \begin{subfigure}
        \centering
        \includegraphics[width=8 cm]
        {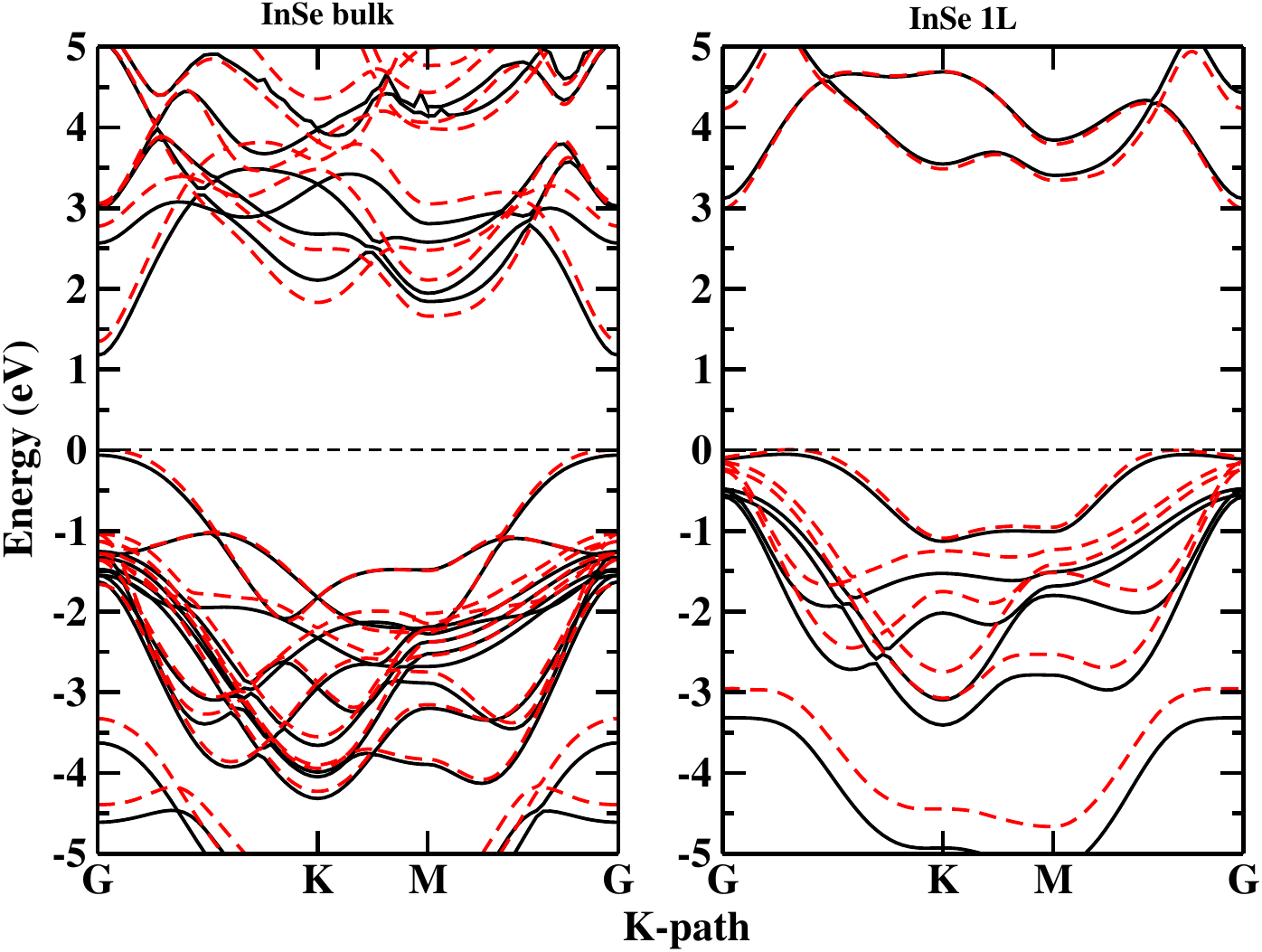}
    \end{subfigure}
    \caption{band structures of MoS$_2$, WS$_2$, $h$BN, BP, and InSe as obtained from SE-DD-RSH and and $G_0W_0$. For bulk BP, $G_0W_0$ is calculated on top of the HSE06. Also, the SE-DD-RSH dielectric constant is calculated using HSE06 for this particular material.}
    \label{band_structure}
\end{figure*}

%

%
%

\subsection{Materials and computational details}

We investigate the electronic and optical properties of nine van der Waals (vdW) materials, 
encompassing bulk and monolayer, from the families of transition metal dichalcogenides (TMDCs), 
hexagonal boron nitride ($h$BN), black phosphorus (BP), and transition metal monochalcogenides (TMMCs). 
Specifically, we examine the trigonal prismatic (2H) phases of MoS$_2$, MoSe$_2$, MoTe$_2$, WS$_2$, WSe$_2$, 
and WTe$_2$, collectively referred to as MX$_2$. Both monolayer and bulk structures are modeled using the 
conventional $AB$ stacking configuration for TMDCs and $AA'$ stacking for $h$BN. Black phosphorus, an 
orthorhombic group-IV monochalcogenide, is also included due to its intriguing anisotropic transport 
properties and strong excitonic effects~\cite{Ling2015renaissance,Tran2014Layer}. We consider both the 
monolayer (phosphorene) and bulk structures, adopting an $AB$ stacking order. Finally, we explore 
$\beta$-InSe as a representative TMMC system, where the group IIIA element (In) forms a hexagonal lattice 
with chalcogen (Se) atoms. A summary of the structural configurations is provided in Fig.~\ref{structure-fig}.


All calculations are performed using the plane-wave pseudopotential code Vienna Ab initio Simulation Package (VASP)~\cite{vasp1,vasp2,vasp3,vasp4}. Structural optimizations are carried out using the 
Perdew-Burke-Ernzerhof (PBE) exchange-correlation functional with the D3(0) dispersion correction. 
The equilibrium volume of bulk structures is first determined at the PBE+D3(0) level, after which the 
optimized in-plane lattice constants are used to construct corresponding monolayer geometries. 
Monolayer structures are further relaxed with fixed out-of-plane parameters to capture intrinsic strain effects.
Electronic band structure calculations are performed on PBE-relaxed geometries using a plane-wave energy 
cutoff of 650 eV. For quasi-particle band gap corrections, we employ the $G_0W_0$ approach with a 
PBE starting point, except for bulk BP, where the HSE06 hybrid functional is used. The macroscopic 
static dielectric tensor $\epsilon_{\infty}^{SC}$ (both in-plane and out-of-plane components) is 
computed within the random phase approximation (RPA) at the PBE level (RPA@PBE), except for bulk BP, 
where RPA@HSE06 is applied. The optical spectra within the Bethe-Salpeter equation (BSE) and 
time-dependent density functional theory (TDDFT) formalisms are computed using a Gaussian broadening 
of $\sigma = 0.1$ eV. The $\mathbf{k}$-point sampling grids used for all calculations are provided 
in Table~\ref{tab-band-gaps}.


Further details on the computational parameters are available in Appendix~\ref{Appendix-B}.
The structural configurations and stacking arrangements of the materials under consideration are illustrated 
in Fig.\ref{structure-fig}. 


\subsection{Layer-dependent band gaps}

We systematically evaluate the performance of various density functional theory (DFT) methods for 
predicting electronic band gaps, as summarized in Table~\ref{tab-band-gaps}. 
Standard functionals such as PBE and HSE06 are included for benchmarking purposes. In the absence of experimental band gap data, single-shot $G_0W_0$ calculations serve as the reference standard 
for assessing the accuracy of different methodologies. 

The performance of the SE-DD-RSH functional is first benchmarked for bulk materials using the effective dielectric 
constant, $\epsilon_\infty^{\text{2D,eff}}$, derived from Eq.~\ref{die-eff1}. For MX$_2$, WX$_2$, and $h$BN, we 
report direct band gaps at the $K$ point, whereas for $\beta$-InSe and black phosphorus (BP), the gaps are reported 
at the $\Gamma$ point. As expected, PBE significantly underestimates the band gaps of all bulk systems. Both 
HSE06 and SE-DD-RSH provide improved predictions for MX$_2$ and WX$_2$, yielding results that closely align with 
$G_0W_0$ calculations.

For the wide-bandgap insulator hexagonal boron nitride ($h$BN), SE-DD-RSH slightly overestimates the band gap 
compared to $G_0W_0$. However, recent studies have emphasized the importance of zero-point renormalization (ZPR) 
corrections in refining band gap estimates~\cite{Camarasa2024excitons}. By incorporating a ZPR correction of 
$\sim 0.3$ eV from Ref.\cite{Camarasa2024excitons}, the SE-DD-RSH band gap is adjusted to approximately 7.30 eV, 
bringing it into closer agreement with values reported at various levels of theory. Furthermore, we explore the 
performance of partially self-consistent $GW$ calculations ($scGW_0$), where the Green’s function $G$ is 
iterated self-consistently while keeping the screened Coulomb interaction $W$ and the orbitals fixed at the PBE level.
As shown in Table~\ref {tab-band-gaps}, the $scGW_0$ band gap for $h$BN is 7.32 eV, demonstrating close agreement 
with SE-DD-RSH and further highlighting the variability introduced by different levels of self-consistency 
in $GW$ methods.

Black phosphorus (BP), on the other hand, is a narrow-bandgap semiconductor with a $G_0W_0$ bandgap of 0.71 eV. 
Due to the metallic character predicted by PBE, $G_0W_0$ calculations are initialized from the HSE06 eigensystem. 
For SE-DD-RSH, the dielectric screening parameter $\epsilon_\infty^{\text{SC}}$ is computed using HSE06. 
We observe that both HSE06 and SE-DD-RSH yield similar band gaps within $\sim 0.2$ eV of $G_0W_0$. 
Similarly, for the 2H phase of $\beta$-InSe, which features a direct gap at $\Gamma$, $G_0W_0$ estimates a 
gap of 1.32 eV. SE-DD-RSH provides an improved prediction of 1.21 eV, outperforming HSE06. 
Notably, for bulk systems, the parameters for SE-DD-RSH are determined in a non-empirical manner, and the method performs very well in predicting band gaps.

Now, we turn to the performance of different methods for monolayers. 
For monolayers of transition-metal dichalcogenides (MoS$_2$, MoSe$_2$, MoTe$_2$, WS$_2$, WSe$_2$, and WTe$_2$),  the PBE functional systematically underestimates the band gaps, with more pronounced errors than in bulk. 
While HSE06 provides an improvement, it still exhibits significant deviations from the precise values. 
In contrast, SE-DD-RSH demonstrates superior performance, closely reproducing $G_0W_0$ results across all 
monolayers (Table~\ref{tab-band-gaps}), confirming that the physically motivated rescaling procedure 
(Eq.\ref{eq-rescale}) and range-separation parameter $\mu$ (Eq.\ref{2d-mu}) appropriately capture the 
screening effects inherent to low-dimensional materials.

In the case of $h$BN, $G_0W_0$ predicts a band gap of 7.42 eV, while PBE and HSE06 significantly underestimate 
this value. SE-DD-RSH provides a band gap of 8.35 eV, which, despite a slight overestimation, aligns closely 
with $scGW_0$. This suggests that the $G_0W_0$ approach, while widely regarded as a state-of-the-art, retains 
some degree of empiricism and sensitivity to methodological approximations. 
Additionally, as noted in Ref.~\cite{Camarasa2024excitons}, incorporating a ZPR correction of 0.38 
eV further improves the agreement between SE-DD-RSH and $G_0W_0$.

For phosphorene, the monolayer of black phosphorus (BP), $G_0W_0$ predicts a band gap of 1.51 eV, with PBE yielding 
an underestimated value of 0.67 eV. SE-DD-RSH provides a highly accurate prediction of 1.48 eV, while HSE06 underestimates the gap at 1.32 eV. 
Similarly, as a second set of examples, monolayer $\beta$-InSe with the 2H structure, $G_0W_0$ estimates a 
band gap of 3.04 eV, while SE-DD-RSH predicts 3.22 eV, demonstrating an excellent agreement. 
In contrast, HSE06 significantly underestimates the gap, predicting a value of 2.49 eV, further highlighting the limitations of fixed hybrid functionals for low-dimensional systems. 



Next, moving to the bilayer structures, a similar trend is observed across different levels of approximation.
SE-DD-RSH consistently outperforms HSE06, providing band gaps in closer agreement with $G_0W_0$ across all 
materials. 
The observed trends reinforce the limitations of HSE06 in describing 2D systems, as its fixed screening parameter 
($\alpha = 0.25$) fails to account for the dimensionality-dependent dielectric 
response~\cite{Patra2020Electronic,patra2021efficient}. This limitation, widely acknowledged in previous 
studies~\cite{Wang2016Hybrid,Tran2021bandgap}, makes HSE06 unsuitable for simultaneously capturing both 
bulk and monolayer properties. In contrast, SE-DD-RSH emerges as a robust alternative, offering improved 
accuracy across different dimensionalities without the need for empirical adjustments. 
For a comprehensive assessment, Fig.~\ref{band_structure} compares the band structures of bulk and monolayer 
MoS$_2$, WS$_2$, $h$BN, BP, and InSe obtained from $G_0W_0$ and SE-DD-RSH. SE-DD-RSH accurately reproduces 
$G_0W_0$ band structures for bulk materials, with the exception of $h$BN. For monolayers, SE-DD-RSH maintains 
strong agreement with $G_0W_0$, further validating its effectiveness in capturing the dielectric screening 
effects that dominate low-dimensional materials. 


\subsection{Linear dielectric response}
\begin{figure*}
    \centering
    \includegraphics[width=17 cm, height=14 cm]{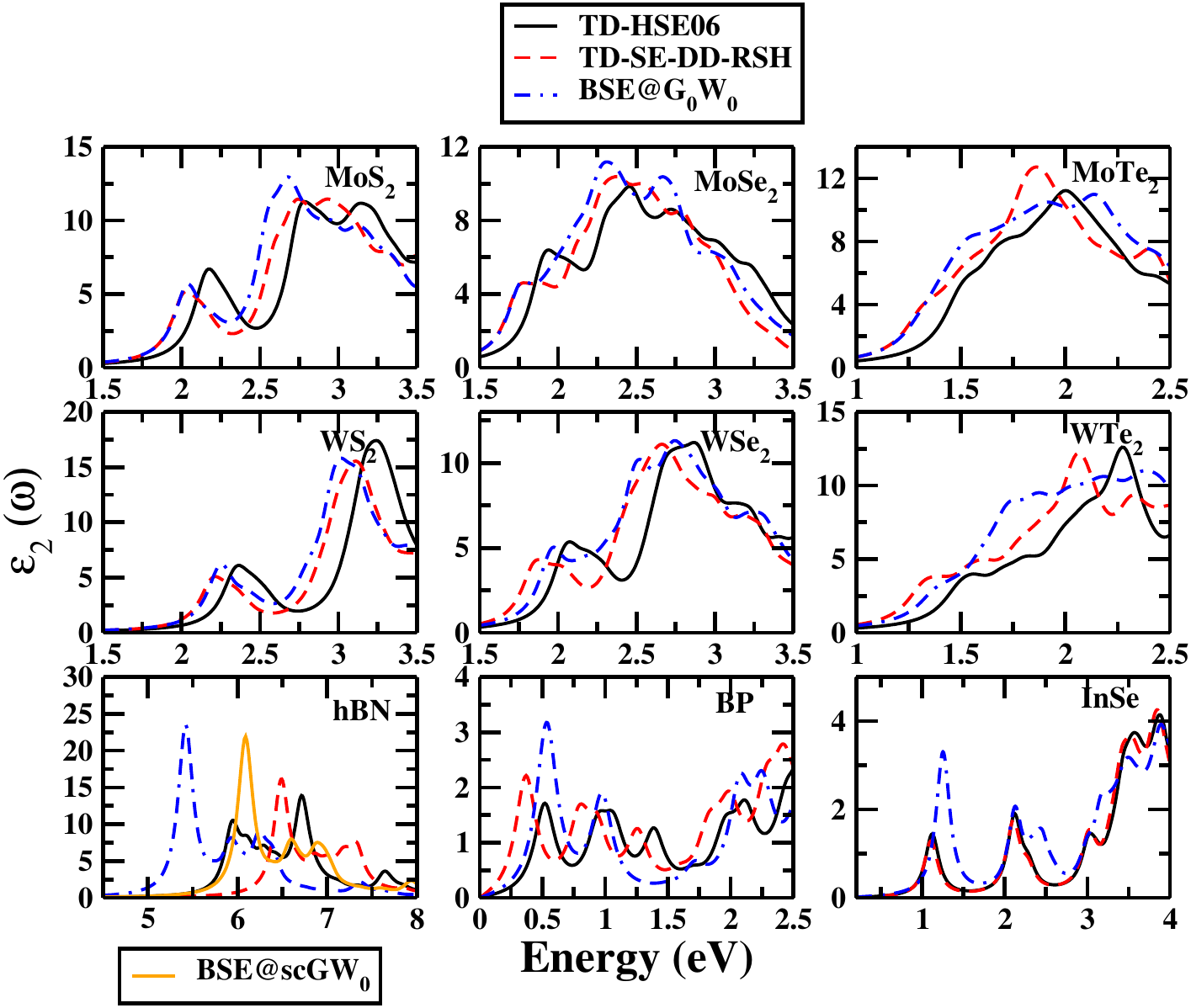}
    \caption{Optical absorption spectra of bulk vdW materials obtained using different methods.}
  \label{bulk-optical-spectra}
\end{figure*}
\begin{figure*}
    \centering
    \includegraphics[width=17 cm, height=14 cm]{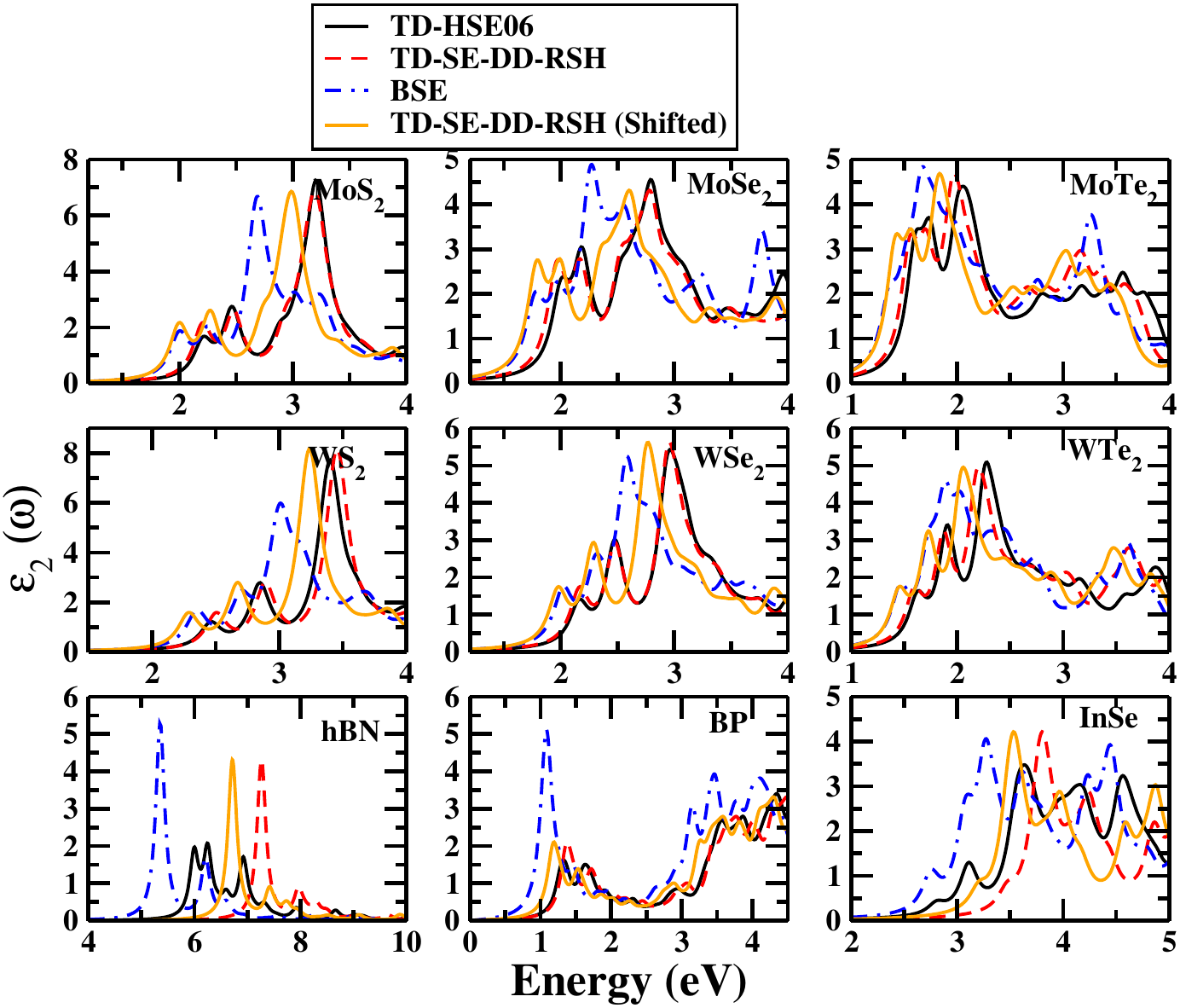}
    \caption{Same as Fig.~\ref{bulk-optical-spectra} but for monolayer vdW materials. The 'shifted' method refers to the empirical shifting on top of the DDH spectra.}
  \label{ml-optical-spectra}
\end{figure*}

\begin{table*}
    \centering
    \renewcommand{\arraystretch}{1.2} 
    \setlength{\tabcolsep}{8pt}      

    \caption{\label{exciton-bulk}  BSE and TDDFT spectral position or excitonic optical transitions and the exciton binding energies $\Delta=E_g^{G_0W_0/DFT}-E^{BSE/TDDFT}$ of the lowest bound excitons of bulk materials. Reported values are without SOC correction. All values are in eV.}
    \begin{tabular}{lcccccccccc}
        \toprule
        \textbf{Solids} & \multicolumn{2}{c}{MBPT} & \multicolumn{2}{c}{HSE06} & \multicolumn{2}{c}{SE-DD-RSH}\\
        \cmidrule(lr){2-3} \cmidrule(lr){4-5} \cmidrule(lr){6-7}
        & BSE  & $\Delta$ & TDDFT  & $\Delta$ & TDDFT  & $\Delta$ \\
        \midrule
        MoS$_2$   & 2.03 & 0.145 & 2.15 & 0.023 & 2.01 & 0.121 \\
        MoSe$_2$  & 1.76 & 0.091 & 1.89 & 0.024 & 1.74 & 0.096 \\
        MoTe$_2$  & 1.30 & 0.071 & 1.48 & 0.022 & 1.30 & 0.071 \\
        WS$_2$    & 2.25 & 0.158 & 2.33 & 0.018 & 2.19 & 0.123 \\
        WSe$_2$   & 1.96 & 0.090 & 2.03 & 0.011 & 1.84 & 0.093 \\
        WTe$_2$   & 1.37 & 0.064 & 1.48 & 0.018 & 1.32 & 0.055 \\
        $h$BN     & 5.35 (6.00$^a$) & 1.291 (1.320$^b$) & 5.88  & 0.547 & 6.40 & 1.185 \\
        BP        & 0.53 & 0.174 & 0.52 & 0.006 & 0.37 & 0.117 \\
        InSe      & 1.25 & 0.093 & 1.12 & 0.010 & 1.10 & 0.140 \\

        \bottomrule
    \end{tabular}
        \footnotetext{partially self-consistent $GW$ ($scGW_0$) calculation.}
        \footnotetext{$\Delta=$ $scGW_0$-BSE@$scGW_0$.}
\end{table*}

\begin{table*}
    \centering
    \renewcommand{\arraystretch}{1.2} 
    \setlength{\tabcolsep}{6pt}      
    \caption{\label{exciton-ml}  BSE and TDDFT spectral position or excitonic optical transitions and the exciton binding energies $\Delta=E_g^{G_0W_0/DFT}-E^{BSE/TDDFT}$ of the lowest bound excitons of monolayers and bilayers. All calculations are without considering SOC.}
    \begin{tabular}{lcccccccccccccc}
        \toprule
        \textbf{Solids} & \multicolumn{2}{c}{MBPT} & \multicolumn{2}{c}{HSE06} & \multicolumn{4}{c}{SE-DD-RSH} & Layers  \\
        \cmidrule(lr){2-3} \cmidrule(lr){4-5} \cmidrule(lr){6-9}
        & BSE & $\Delta$ & TDDFT& $\Delta$ & TDDFT & $\Delta$& TDDFT+Shift&(TDDFT+Shift)$-$BSE  \\
        \midrule
MoS$_2$	&	2.00	&	0.630	&	2.21	&	0.029	&	2.20	&	0.188	&	2.00	&	-0.002	&	1L	\\
	&	2.01	&	0.397	&	2.18	&	0.011	&	2.17	&	0.154	&	1.98	&	-0.032	&	2L	\\
MoSe$_2$	&	1.77	&	0.564	&	2.00	&	0.037	&	1.97	&	0.169	&	1.79	&	0.011	&	1L	\\
	&	1.78	&	0.344	&	1.98	&	0.014	&	1.93	&	0.131	&	1.77	&	-0.009	&	2L	\\
MoTe$_2$	&	1.36	&	0.457	&	1.60	&	0.042	&	1.55	&	0.123	&	1.41	&	0.056	&	1L	\\
	&	1.32	&	0.289	&	1.55	&	0.026	&	1.50	&	0.087	&	1.37	&	0.051	&	2L	\\
WS$_2$	&	2.35	&	0.643	&	2.46	&	0.015	&	2.51	&	0.176	&	2.29	&	-0.065	&	1L	\\
	&	2.34	&	0.413	&	2.43	&	0.002	&	2.46	&	0.153	&	2.25	&	-0.088	&	2L	\\
WSe$_2$	&	2.02	&	0.572	&	2.16	&	0.017	&	2.17	&	0.150	&	1.98	&	-0.045	&	1L	\\
	&	1.99	&	0.382	&	2.12	&	0.004	&	2.12	&	0.120	&	1.94	&	-0.056	&	2L	\\
WTe$_2$	&	1.46	&	0.458	&	1.62	&	0.023	&	1.59	&	0.100	&	1.45	&	-0.011	&	1L	\\
	&	1.41	&	0.295	&	1.57	&	0.003	&	1.53	&	0.081	&	1.39	&	-0.018	&	2L	\\
$h$BN	&	5.35	&	2.064	&	5.99	&	0.111	&	7.26	&	1.090	&	6.71	&	1.363	&	1L	\\
	&	5.33	&	1.767	&	6.01	&	0.148	&	7.21	&	1.168	&	6.67	&	1.342	&	2L	\\
BP	&	1.08	&	0.428	&	1.29	&	0.031	&	1.37	&	0.118	&1.18		&	0.097	&	1L	\\
	&	0.64	&	0.245	&	0.85	&	0.028	&	0.82	&	0.074	&	0.68	&	0.040	&	2L	\\
InSe$^a$	&	2.68	&	0.403	&	2.53	&	0.003	&	3.11	&	0.158	&	2.84	&	0.158	&	1L	\\
	&	2.02	&	0.259	&	1.90	&	-0.003	&2.42		&0.107		&2.16		&0.147		&	2L	\\
        \bottomrule
    \end{tabular}
            \footnotetext{Experimental excition peak at 2.92 eV~\cite{Guo2017Band}.}
\end{table*}

\begin{table*}
    \centering
\renewcommand{\arraystretch}{1.2} 
    \setlength{\tabcolsep}{8pt}      

    \caption{\label{monolayer-expt-comparison}  Comparison of band gaps and binding energies of theoretically obtained values (without SOC corrected) with experimental results. The experimental results are taken from Table II of ref.~\cite{Wang2018RPM}.}
    \begin{tabular}{lcccccccccccccccc}
        \toprule
        \textbf{Solids} & \multicolumn{2}{c}{MBPT} & \multicolumn{2}{c}{SE-DD-RSH} & \multicolumn{2}{c}{Expt.}\\
        \cmidrule(lr){2-3} \cmidrule(lr){4-5} \cmidrule(lr){6-7}
        & $E_g$ & BE & $E_g$ & BE& $E_g$ & BE \\
        \midrule
        MoS$_2$   & 2.63&0.63&2.39&0.188&2.15$\pm$0.06&0.2 (or 0.42)  \\
        &&&&&2.5&$\geq$0.57\\
        &&&&&2.47$\pm$0.08&0.44$\pm$0.08\\
        &&&&&2.15$\pm$0.1&0.3\\
        &&&&&2.17$\pm$0.1&0.31$\pm$0.04\\[0.5 cm]
        MoSe$_2$  &2.34&0.564&2.14&0.169&2.18&0.55  \\
        &&&&&2.15$\pm$0.06&0.5\\[0.5 cm]
        WS$_2$    &2.99&0.643&2.68&0.176&2.41$\pm$0.04&0.32$\pm$0.04  \\
        &&&&&2.7&0.7\\
        &&&&&2.73&0.71$\pm$0.01\\
        &&&&&3.01&0.929\\
        &&&&&2.33$\pm$0.05&0.32$\pm$0.05\\
        &&&&&2.38$\pm$0.06&0.36$\pm$0.06\\
        &&&&&2.31$-$2 .53&0.26 - 0.48\\[0.5 cm]
        WSe$_2$   &2.59&0.572&2.31&0.150&2.02&0.37  \\
        &&&&&2.2$\pm$0.1&0.5\\
        &&&&&2.35$\pm$0.2&0.6$\pm$0.2\\
        &&&&&2.63&0.887\\
        &&&&&2.08$\pm$0.1&0.4\\
        &&&&&1.9&0.245\\
        \bottomrule
    \end{tabular}
\end{table*}

Next, we investigate the dielectric response and optical absorption spectra of bulk, monolayer, and bilayer 
structures of MX$_2$, WX$_2$, $h$BN, BP, and InSe, employing time-dependent SE-DD-RSH (TD-SE-DD-RSH) 
and benchmark our findings against Bethe-Salpeter equation calculations within the BSE@$G_0W_0$.
The optical excitations are computed using linear-response time-dependent density functional theory (LR-TDDFT) by solving the Casida/BSE equation~\cite{cassida-equation} with generalized Kohn-Sham (gKS) orbitals and eigenvalues.
Within the Tamm-Dancoff approximation, transition matrix elements are systematically evaluated following Refs.~\cite{cassida-equation,Sander2015beyond,Sander2017macroscopic,tal2020accurate}, ensuring a rigorous description of many-body interactions and excitonic effects across different dimensionalities.
%
We consider the excitonic Hamiltonian,
\begin{eqnarray}
A_{ai;bj} = \omega_{ai} \delta_{ij;ab} + \langle ib | \hat{K} | aj \rangle
\end{eqnarray}
where $\omega_{ai}=\varepsilon_a-\varepsilon_i$ represents the energy difference between unoccupied ($a, b$) 
and occupied ($i,j$) are the occupied states within the gKS framework. The interaction term 
$\langle ib|\hat{K}|aj\rangle$ determines the nature of excitonic effects and takes distinct forms 
within TDDFT and BSE: 
\begin{eqnarray}
 \langle ib|\hat{K}|aj\rangle &=&  2\langle ib|\hat{V}_H|aj\rangle 
 - \langle ib|\hat{W}|ja\rangle~, 
 \label{eq-K}
\end{eqnarray}
where the Hartree term $V_H$ is universal. 
For TDDFT, the screened exchange interaction 
is expressed as
\begin{eqnarray}
 \langle ib|W|ja\rangle &=&  \langle ib|f_{xc}^{non-local}|ja\rangle-\langle ib|f_{xc}^{local}|aj\rangle
\end{eqnarray}
with non-local XC kernel, $f_{xc}^{non-local}$ becomes Fock exchange term screened by dielectric function, $[\epsilon_\infty^{2D,eff}]^{-1}$ i.e., $\langle ib|f_{xc}^{non-local}|ja\rangle = \langle ib|[\epsilon_\infty^{2D,eff}]^{-1}\hat{V}_{Fock}|ja\rangle$ and the local XC kernel in TD-DFT calculation is given by $\langle ib|f_{xc}^{local}|aj\rangle$, where
\begin{equation}
f_{xc}^{local}({\bf{r}},{\bf{r}}')=\frac{\delta^2[E_c^{\text{PBE}}+(1-[\epsilon_\infty^{\text{2D,eff}}]^{-1})E_x^{\text{PBE}}]}{\delta n({\bf{r}})\delta n({\bf{r}'})}
\end{equation}
%

In contrast, BSE incorporates a dynamically screened interaction $\hat{W}(\omega)$, making the resulting excitation 
spectra more physically accurate. However, due to its computational expense, dynamical effects are often approximated 
or neglected.

Finally, the optical response within TDDFT and BSE is obtained from the imaginary part of the macroscopic 
dielectric function
$\epsilon_2(\omega)$ 
as~\cite{Sander2017macroscopic}, 
\begin{eqnarray}
\epsilon_2(\omega) = \Im \left\{ \lim_{q \rightarrow 0} \epsilon^M(q, \omega) \right\}~,
\label{eq8bb}
\end{eqnarray}
%

Note that for layer materials, $\epsilon_2(\omega)$ is averaged over the in-plane components 
$\epsilon_2(\omega)=\frac{\epsilon_{xx}(\omega)+\epsilon_{yy}(\omega)}{2}$, 
while for bulk systems, a three-dimensional average is taken 
$\epsilon_2(\omega)=\frac{\epsilon_{xx}(\omega)+\epsilon_{yy}(\omega)+\epsilon_{zz}(\omega)}{3}$. 
For computational feasibility, spin-orbit coupling (SOC) effects are neglected, though their influence is minor.
Additionally, Table~\ref{exciton-bulk} and Table~\ref{exciton-ml}, we present the peak positions obtained 
from TDDFT and BSE calculations, along with their respective deviations from the corresponding direct band gaps.

Figure~\ref{bulk-optical-spectra} presents the optical absorption spectra for bulk van der Waals materials, revealing the efficacy of the TD-SE-DD-RSH approach in accurately capturing excitonic effects. As illustrated in both Fig.\ref{bulk-optical-spectra} and Table\ref{exciton-bulk}, the optical response predicted by TD-SE-DD-RSH exhibits remarkable agreement with the many-body perturbation theory (MBPT) results obtained using the Bethe-Salpeter equation atop $G_0W_0$ (BSE@$G_0W_0$). For instance, in the case of MoS$_2$, the TD-SE-DD-RSH optical gap, corresponding to the first excitonic peak, is computed at 2.01 eV, in excellent agreement with the BSE@$G_0W_0$ result of 2.03 eV. Similarly, for MoSe$_2$ and MoTe$_2$, the peak positions predicted by TD-SE-DD-RSH are 1.74 eV and 1.30 eV, respectively, closely matching the corresponding BSE@$G_0W_0$ values of 1.76 eV and 1.30 eV. The deviations remain minimal across the transition-metal dichalcogenide (TMDC) family, with WS$_2$ and WSe$_2$ exhibiting discrepancies of merely $\sim$0.06 eV and $\sim$0.03 eV, respectively. This consistency extends to WTe$_2$, where the peak position deviation remains within $\sim$0.09 eV. From a theoretical standpoint, the exciton binding energy, defined as the difference between the optical excitation energy (TDDFT or BSE) and the quasiparticle band gap (or DFT)—emerges as a critical benchmark for validating the accuracy of computational approaches. Our analysis underscores the striking agreement between BSE@$G_0W_0$ and TD-SE-DD-RSH, reinforcing the latter's ability to encapsulate the intricate electron-hole interactions that govern the optical response of layered materials. In contrast, spectra computed using TD-HSE06 consistently deviate from BSE@$G_0W_0$ and TD-SE-DD-RSH, particularly underestimating exciton binding energies. This systematic discrepancy arises from the absence of long-range screened exchange in HSE06, leading to an incorrect asymptotic behavior of the exchange potential.

A notable exception to the observed agreement is found in hexagonal boron nitride ($h$BN), where the TD-SE-DD-RSH exciton peak appears at 6.40 eV—approximately 1 eV higher than the BSE@$G_0W_0$ value. This deviation is indicative of the well-documented challenges associated with accurately predicting the band gap of wide-gap insulators within conventional electronic structure formalisms~\cite{Camarasa2024excitons}. However, previous theoretical investigations~\cite{Camarasa2024excitons} report exciton peak energies in the range of 6.00 eV to 6.52 eV, depending on the self-consistency level of the underlying MBPT treatment. Additionally, the importance of ZPR effects in $h$BN has been emphasized, with an estimated correction of $\sim$0.3 eV~\cite{Camarasa2024excitons}. Incorporating this ZPR correction into our TD-SE-DD-RSH result yields a renormalized excitonic peak at $\sim$6.10 eV, bringing it closer to experimental measurements (5.99 eV)\cite{Tarrio1989Interband} and advanced self-consistent many-body calculations such as $scGW_0$ (Table\ref{exciton-bulk}).

A similar assessment for black phosphorus (BP) reveals a systematic deviation of $\sim$0.2 eV between TD-SE-DD-RSH and BSE@$G_0W_0$, further confirming the former's reliability in describing excitonic effects in anisotropic layered semiconductors. Notably, for InSe, the first optical peak exhibits a marginal $\sim$0.1 eV discrepancy between TD-SE-DD-RSH and BSE, emphasizing the robustness of the dielectric-dependent hybrid approach across different material classes.

Overall, while the HSE06 hybrid functional provides a reasonable estimate for quasiparticle band gaps, it systematically underestimates exciton binding energies due to the absence of nonlocal dielectric screening effects. This behavior is inherently linked to its formulation, which includes only short-range Hartree-Fock exchange, thereby failing to reproduce the correct asymptotic behavior of the screened exchange potential. 

Regarding the optical absorption spectrum of monolayers, Fig.~\ref{ml-optical-spectra} provides a comparative analysis of various theoretical approaches in capturing excitonic effects. The spectral positions and exciton binding energies of the lowest bound excitons, summarized in Table~\ref{exciton-ml}, reveal the profound impact of reduced Coulomb screening in low-dimensional systems. Unlike their bulk counterparts, monolayer materials exhibit significantly enhanced exciton binding energies due to the diminished dielectric screening, leading to stronger electron-hole interactions and more localized excitonic wavefunctions. Within this framework, the BSE@$G_0W_0$ formalism, which fully accounts for the frequency-dependent screening effects, remains the gold standard for computing exciton binding energies. Our analysis confirms that TD-SE-DD-RSH systematically approximates these values with high accuracy, whereas TD-HSE06 considerably underestimates exciton binding energies, often approaching zero. This discrepancy underscores the fundamental limitation of the HSE06 functional, which lacks the necessary long-range screened exchange to accurately describe excitonic states in two-dimensional (2D) materials. In contrast, TD-SE-DD-RSH provides a substantial improvement over TD-HSE06, capturing the essential screening physics while maintaining computational efficiency. For instance, in monolayer MoS$_2$, the spectral positions predicted by TD-SE-DD-RSH are systematically blue-shifted by approximately 100 meV relative to TD-HSE06. This trend is consistently observed across other chalcogenide materials, reaffirming the robustness of the TD-SE-DD-RSH framework in 2D systems. Notably, for all TMDCs, the first bright excitonic peak is located at the K point of the Brillouin zone, in agreement with both theoretical and experimental studies.

Turning to the monolayer of hexagonal boron nitride ($h$BN), we find that the first excitonic peak position and corresponding exciton binding energy within TD-SE-DD-RSH are 7.26 eV and 1.09 eV, respectively. These values align closely with BSE@$GW_0$ results, highlighting the accuracy of our approach in capturing strong excitonic effects in wide-gap insulators. As discussed in prior theoretical works~\cite{Ramasubramaniam2019Transferable, Camarasa2023Transferable}, the determination of excitonic energies in $h$BN remains particularly challenging due to the interplay of electronic correlation and ZPR effects. Recent studies~\cite{Camarasa2024excitons} suggest that an additional ZPR correction may further refine the agreement with experimental measurements, underscoring the necessity of incorporating phonon-driven renormalization effects in future studies.

For the BP monolayer, our calculations indicate that the first excitonic peak obtained from BSE occurs at 1.08 eV, which is approximately 0.3 eV lower than the TD-SE-DD-RSH value at the $\Gamma$ point. The corresponding exciton binding energies exhibit similar trends, with TD-SE-DD-RSH providing a significantly more accurate description than TD-HSE06. A similar performance is observed for monolayer InSe, where the first excitonic peak position obtained from TD-SE-RS-DDH (3.11 eV) is in close agreement with BSE@$GW$ (2.68 eV). This prediction also aligns well with the experimentally measured optical gap of 2.92 eV~\cite{Guo2017Band}, reinforcing the validity of TD-SE-RS-DDH as a computationally efficient alternative to many-body perturbation theory.

In principle, these discrepancies in linear optical absorption spectra between BSE and TDDFT arise primarily from differences in their treatment of the screening effects. Unlike $G_0W_0$, the conventional DFT and TDDFT screened exchange kernel lacks frequency dependence, leading to systematic underestimations of exciton binding energies. To mitigate these discrepancies and provide a more realistic description of exciton peak positions, and hence the binding energies, in the following we propose a dielectric-dependent shited procedure to the TD-SE-DD-RSH optical spectra: 
\begin{eqnarray}
E^{TDDFT-Shifted}&=&E^{TDDFT}-a[\epsilon_\infty^{2D}]^{-1}\\
E_b^{TDDFT-Shifted}&=&E_g^{SE-DD-RSH}-E^{TDDFT-Shifted}\nonumber\\
&=&E_b^{TDDFT}+a[\epsilon_\infty^{2D}]^{-1}
\end{eqnarray}
with $a$ being the fitting parameter determined by comparison with BSE results, with an optimal value of 1.8 eV. The effectiveness of this shifted-TD-SE-DD-RSH method is demonstrated in Fig.~\ref{ml-optical-spectra} and Table~\ref{exciton-ml}, where the corrected exciton peak positions and binding energies exhibit remarkable agreement with BSE@$G_0W_0$. Conceptually, this correction is analogous to the scissor operator approach commonly employed in quasiparticle band gap corrections, where the difference between the quasiparticle and Kohn-Sham band gaps dictates the magnitude of the shift.  

Finally, in Table~\ref{monolayer-expt-comparison}, we compare the theoretical values with available experimental results. The experimental data were obtained at different temperatures and on various exfoliated surfaces. Despite these variations, the trends indicate that both the $G_0W_0$ and SE-DD-RSH band gaps fall within the range of the experimental values. However, the binding energies obtained from TD-SE-DD-RSH are underestimated, although, for some systems, they remain within the lower limit of the experimental range.

\begin{figure*}
    \centering
    \includegraphics[width=18 cm, height=11 cm]{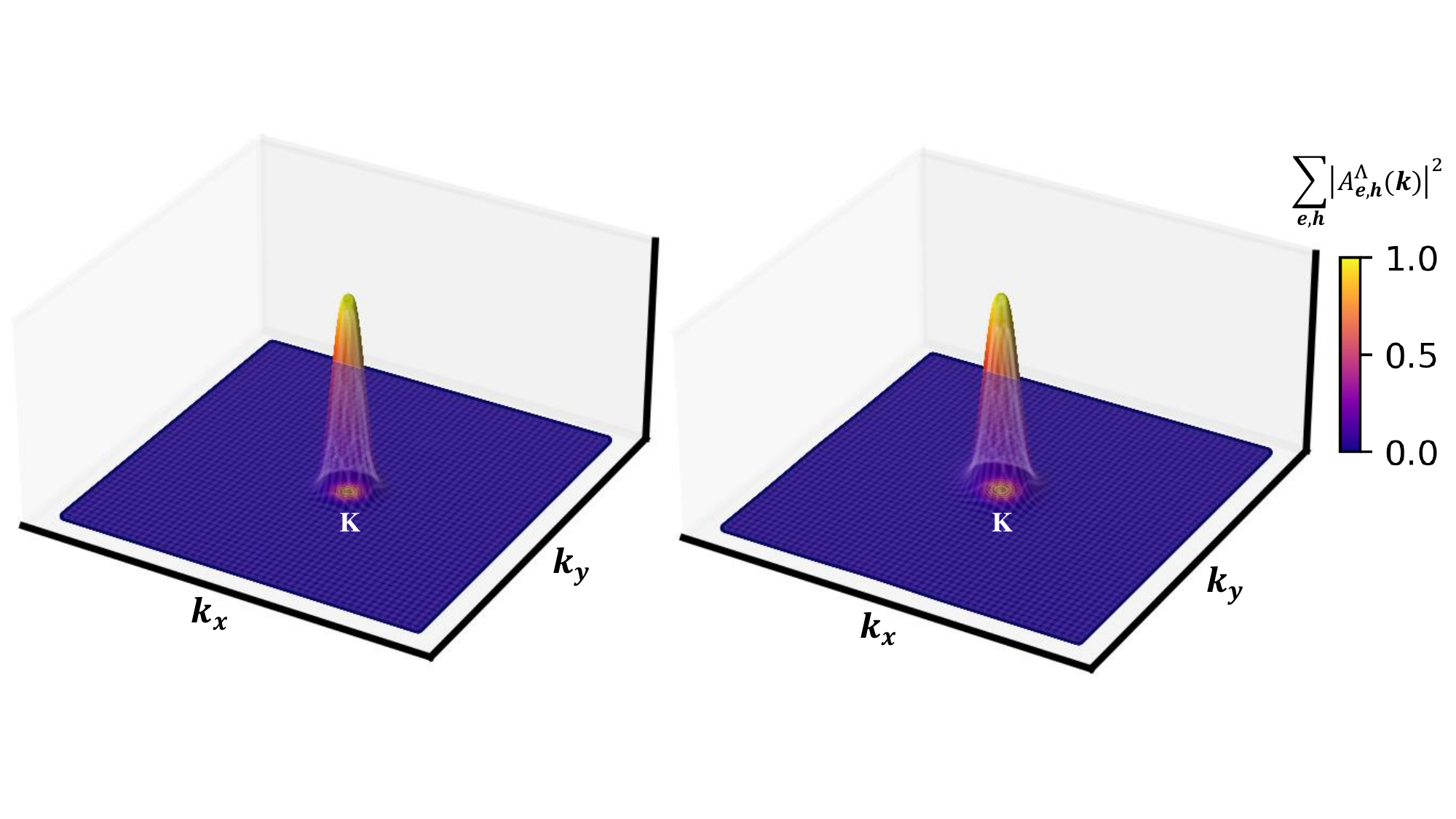}
    \caption{Squared modulus amplitude $|A_{e,h}^{\Lambda}({\bf{k}})|^2$ of the first bright exciton for MoS$_2$ monolayer along the $k_x-k_y$ plane ($k_z = 0$) in the Brillouin zone. The left panel is for BSE and the right panel is for TD-SE-DD-RSH. The eigenvectors are extracted with the $24\times 24\times 1$ ${\bf{k}}-$point for both calculations.}
  \label{exciton-plot}
\end{figure*}

High exciton binding energies from BSE for monolayers and bilayers indicate that its wavefunction is more localized than TD-SE-DD-CAM. In order to understand this in Fig.~\ref{exciton-plot} we plot  squared modulus of exciton amplitude $|\sum_{e,h} A_{e,h}^{\Lambda}({\bf{k}})|^2$ for $e-h$ pair with $\Lambda=1$ i.e., first bright exciton. The size of the projected circle denotes the
absolute amplitude. As shown in Fig.~\ref{exciton-plot}, for this particular example of MoS$_2$ monolayer the BSE amplitude becomes localized than TD-SE-DD-RSH at $K-$point. Interestingly, a close resemblance between both squared amplitudes indicates the closeness of both spectra, although there is a difference of 0.4 eV in exciton binding energies. In this plot, both TDDFT and BSE spectra are produced with a very dense regular $\bf{k}$ grid $24\times 24\times 1$, and we use Gaussian broadening~\cite{Wang2017Electrinic},
\begin{equation}
A_{e,h}^{\Lambda}({\bf{k}})=\sum_{k'}\frac{A_{e,h}^{\Lambda}({\bf{k}})w_{k'}}{(\sigma\sqrt{2\pi})^3} \exp(-\frac{|\bf{k}-\bf{k'}|^2}{2\sigma^2})   
\end{equation}
with $w_{k'}$ being the weight and $\sigma=0.05$ being the broadening.

Also, it is well known that the imaginary part of the dielectric response is related to the exciton amplitude via   
\begin{eqnarray}
    \epsilon_2(\omega) \propto 
    \sum_{\Lambda} \underbrace{\sum_{e,h} \left| A_{e,h}^{\Lambda}({\bf{k}}) \langle e|\vec{D}|h \rangle \right|^2}_{\sum_{eh} |A_{e,h}^{\Lambda}({\bf{k}})|^2} \delta(E_e-E_h- \omega)~,
\end{eqnarray}
where $\vec{D}$ is the transition dipole moment.

%
%



\subsection{MoS$_2$/WS$_2$ heterostructure}

\begin{table*}
    \centering
    \renewcommand{\arraystretch}{1.2} 
    \setlength{\tabcolsep}{8pt}      

    \caption{\label{hetero} DFT ($G_0W_0$) and TDDFT (BSE) band gaps and excitonic positions of the lowest bound excitons for MoS$_2$/WS$_2$ heterostructure. Reported values are without SOC correction. All values are in eV.}
    
    \begin{tabular}{lccccccccc}
        \toprule
        \multicolumn{3}{c}{MBPT} & \multicolumn{3}{c}{HSE06} & \multicolumn{3}{c}{SE-DD-RSH$^a$}&Expt. Peak \\
        \cmidrule(lr){1-3} \cmidrule(lr){4-6} \cmidrule(lr){7-9}
        $G_0W_0$ & BSE & $\Delta$ & $E_g$ & TDDFT & $\Delta$ & $E_g$ & TDDFT & $\Delta$ \\
        \midrule
2.33&1.91&0.425&1.92&1.95&0.011&2.08&1.90&0.129&1.882~\cite{Kaplan2016Excitation}\\        
        \bottomrule
    \end{tabular}
            \footnotetext{A shift of 0.212 eV yields a TDDFT+Shift value of 1.74 eV, with the corresponding binding energy ($\Delta$) being 0.341 eV.}    
\end{table*}

\begin{figure*}
    \centering
    \includegraphics[width=17 cm, height=12 cm]{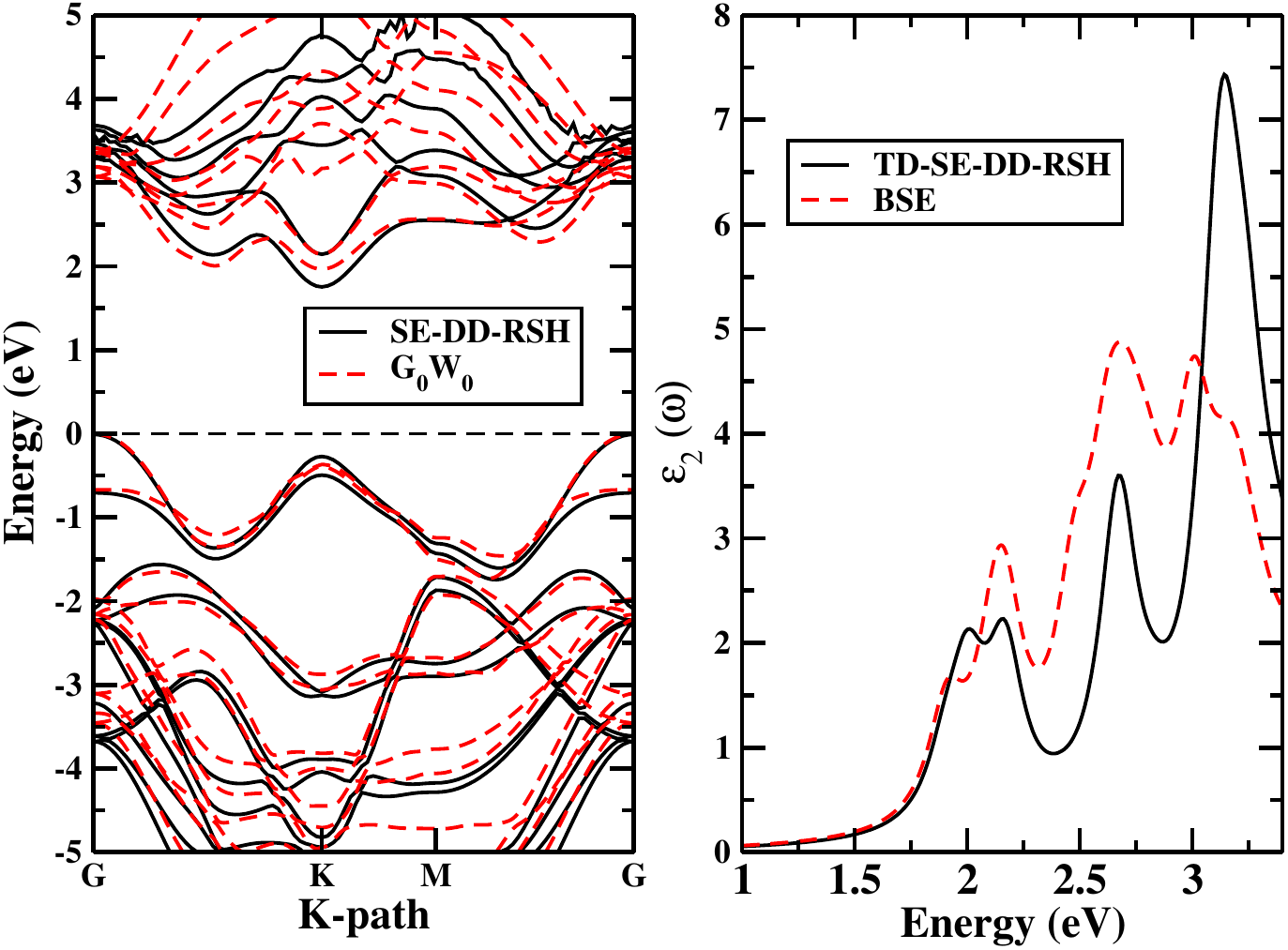}
    \caption{Band structure and optical absorption spectra of MoS$_2$/WS$_2$ heterostructure.}
  \label{hetero-band-optical}
\end{figure*}

To rigorously evaluate the transferability and predictive accuracy of the screened-exchange dielectric-dependent range-separated hybrid (SE-DD-RSH) functional, we extend its application to the MoS$_2$/WS$_2$ heterostructure in the $AA'$ stacking configuration, focusing on the electronic band structure, optical absorption spectra, and excitonic states. The atomic structure of the heterobilayer is adopted from the optimized configuration reported in Ref.~\cite{Torun2018Interlayer}, with the in-plane lattice parameter constrained to the experimental bulk value of 3.162 \AA.

The band structures computed using $G_0W_0$@PBE and SE-DD-RSH are presented in Fig.~\ref{hetero-band-optical}, along with the optical absorption spectra obtained from BSE@$G_0W_0$ and TD-SE-DD-RSH. Notably, spin-orbit coupling (SOC) effects are neglected in these calculations, following the discussions in Ref.~\cite{Torun2018Interlayer}. The direct quasiparticle band gap at the K point, as obtained from $G_0W_0$, is 2.33 eV, serving as a reference for benchmarking the performance of hybrid functionals.

The SE-DD-RSH calculations, we employ $[\epsilon^{2D}_\infty]^{-1}=0.1180$ and $\mu_{layer}=1.096$ \AA$^{-1}$, ensuring consistency with the underlying dielectric environment. The out-of-plane lattice parameter ($c=40.0$ \AA) and the effective layer thickness ($t=12.37118$ \AA) are derived from structural relaxations, providing the necessary input for computing $[\epsilon^{2D}_\infty]^{-1}$. 
The definition of heterobilayer thickness ($t$) plays a crucial role in normalizing the in-plane and out-of-plane dielectric screening properties. We adopt a systematic scaling framework, wherein the thickness of the heterostructure is computed as: $t=\frac{t_{MoS_2}+t_{WS_2}}{2}+d_{interlayer}$, 
where, $t_{MoS_2}$ and $t_{WS_2}$ represent the monolayer thicknesses of MoS$_2$ and WS$_2$, respectively (as determined from their corresponding bilayer structures), while $d_{interlayer}$ denotes the equilibrium interlayer spacing in the heterobilayer configuration. This approach establishes a consistent framework for defining the hetero-bilayer thickness, providing a reliable basis for normalizing the in-plane and out-of-plane dielectric constants using the established scaling relations. On the other hand, for $\mu_{layer}$, we consider the same formula as Eq.~\ref{2d-mu} with $N_e=6$ (the same valance electron for Mo or W). Finally, the band gap obtained from SE-DD-RSH is 2.08 eV, in close agreement with the $G_0W_0$ result.  In comparison, HSE06 significantly underestimates the band gap, yielding a direct transition of 1.92 eV, further reinforcing the necessity of incorporating dielectric-dependent screening in hybrid functionals.

For the optical spectra, the first bright interlayer excitation at the $K$ point is found at 1.91 eV using BSE@$G_0W_0$, resulting in an exciton binding energy of 0.425 eV. In the case of TD-SE-DD-RSH, the first bright excitonic state appears at 1.90 eV, with a corresponding exciton binding energy of 0.129 eV. However, applying an empirical shift on top of TD-SE-DD-RSH shifts the first peak to 0.212 eV, corresponding to a binding energy of 0.341 eV. Overall, the results obtained from TD-SE-DD-RSH and its shifted version are quite encouraging and can be further used as benchmarks for future studies on the excitonic effects of the hetero-bilayer materials. One may note that the experimental exciton peak is obtained to be at 1.882 eV~\cite{Kaplan2016Excitation}. The consistency between our theoretical predictions and experimental findings underscores the physical robustness of the SE-DD-RSH approach in capturing excitonic effects in van der Waals heterostructures.

\section{Conclusions}
\label{conclusion}

The screened-exchange dielectric-dependent range-separated hybrid (SE-DD-RSH) approach presents a computationally efficient and physically robust alternative to many-body perturbation theory ($GW$) calculations. Unlike conventional hybrid functionals that rely on a fixed fraction of Fock exchange, SE-DD-RSH dynamically incorporates system-dependent dielectric screening, making it a more adaptive and physically motivated framework for electronic structure calculations.

In this work, we have developed a first-principles framework for constructing system-specific screening parameters, enabling the accurate modeling of quasiparticle band structures and optical absorption spectra in bulk, two-dimensional (2D), and heterostructured van der Waals (vdW) materials. Our TD-SE-DD-RSH formalism, applied to transition metal dichalcogenides (TMDCs), transition metal monochalcogenides (TMMCs), $\beta$-InSe, and black phosphorus (BP), demonstrates excellent agreement with $G_0W_0$ quasiparticle energies and BSE@$G_0W_0$ optical spectra, establishing its reliability in describing excitonic effects across different material classes.

A key strength of this approach is its formal consistency with dielectric-dependent hybrid (DDH) functionals developed for bulk systems, where the screening parameters are directly obtained from the dielectric response of the material. This feature makes SE-DD-RSH highly transferable, allowing it to be systematically applied to a broad range of low-dimensional quantum materials without requiring empirical tuning from higher-level methods.

Overall, the SE-DD-RSH functional and its time-dependent framework outperform HSE06 in predicting band gaps and excitonic properties. This functional can be recommended as it also employs minimal empiricism. Beyond the systems studied here, this framework can be extended to emerging research directions, including vdW spin valves\cite{Cardoso2018VanderWaals}, vdW heterostructures with moiré superlattices\cite{Geim2013heterostructures}, and 2D magnetic semiconductors~\cite{Li2022Emergent}. The ability of SE-DD-RSH to capture complex many-body interactions while maintaining computational efficiency paves the way for predictive modeling of excitonic phenomena, strongly correlated states, and next-generation optoelectronic devices in layered materials. 



\section*{Acknowledgements}
S.\'S. acknowledges the financial support from the National Science Centre, Poland (grant no. 2021/42/E/ST4/00096)

\section*{Data availability}
All data supporting this study are available within the paper. The structural files used in this work are openly accessible on Zenodo~\cite{zenodo1}. Additional data are available from the authors upon reasonable request.

\appendix

\section{Proof of Eq.~\ref{eq-rescale} and Eq.~\ref{die-eff2}}
\label{Appendix-A}

The proof of Eq.~\ref{eq-rescale} is based on the concept of the two dielectric media arranged parallel and perpendicular to the applied electric field with thickness $t$ and supercell thickness $c$. This is shown in Fig.~\ref{dielectric}. The thickness, $t$ is calculated by computing $n-$ layer thickness considering the structure of $n+1$ -layer as discussed in ref.~\cite{Laturia2018dielectric}.

\setcounter{figure}{0}
\renewcommand{\thefigure}{A\arabic{figure}}
\begin{figure}
    \centering
    \includegraphics[width=\columnwidth]{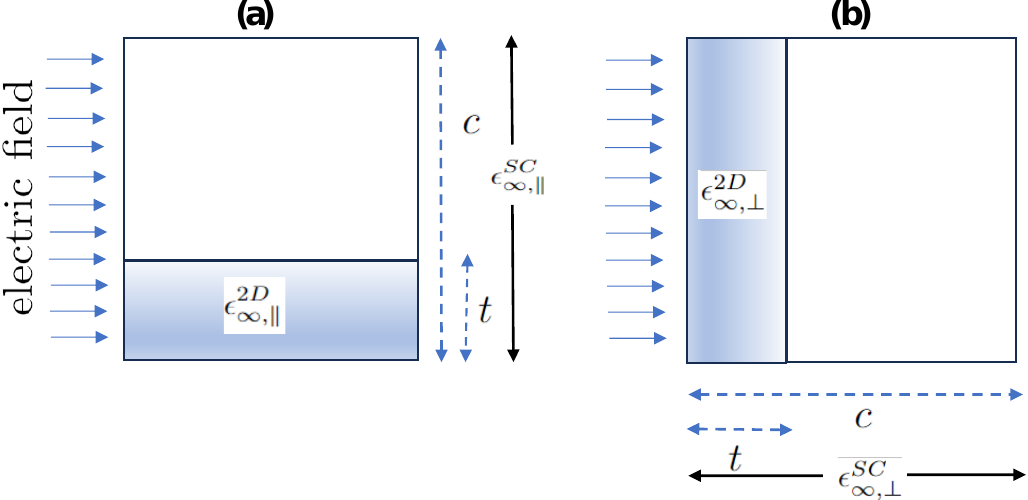}
    \caption{Schematic diagram of two dielectric mediums (2D materials and vacuum) with the plane (a) parallel and (b) perpendicular to the applied electric field.}
  \label{dielectric}
\end{figure}

When the two dielectric mediums, such as 2D material and vacuum, are arranged
 $\parallel$ to the applied electric field, the dielectric constant of the supercell (SC), consisting of the 2D material and vacuum, is given by
\begin{equation}
c~\epsilon^{SC}_{\infty,\parallel}=t~\epsilon^{2D}_{\infty,\parallel}+(c-t)~.
\label{appendix-eq1}
\end{equation}
Here, we consider the dielectric constant of the vacuum to be $\epsilon^{vaccum}_{\infty}=1$.
Using Eq.~\ref{appendix-eq1}, we readily obtain the first sets of Eq.~\ref{eq-rescale}, i.e.,
\begin{eqnarray}
\epsilon^{2D}_{\infty\parallel}&=&1+\frac{c}{t}(\epsilon^{SC}_{\infty\parallel}-1)~.
\label{eq-rescale1}
\end{eqnarray}
Similarly, when the dielectric mediums are arranged $\perp$ to the applied electric field, the dielectric constant of the supercell (SC), consisting of the 2D material and vacuum, is obtained as~\cite{Aspnes1982local},
\begin{equation}
\frac{c}{\epsilon^{SC}_{\infty,\perp}}=\frac{t}{\epsilon^{2D}_{\infty,\perp}}+(c-t)~.
\label{appendix-eq2}
\end{equation}
From Eq.~\ref{appendix-eq2}, we obtain the second set of Eq.~\ref{eq-rescale}, i.e.,
\begin{eqnarray}
\epsilon^{2D}_{\infty\perp}&=&[1+\frac{c}{t}(\frac{1}{\epsilon^{SC}_{\infty\perp}}-1)]^{-1}~.
\label{eq-rescale2}
\end{eqnarray}

Now in order to proof Eq.~\ref{die-eff1}, i.e., 
\begin{equation}
\epsilon^{\text{2D, eff}}_\infty = \sqrt{\epsilon_{\infty\parallel}^{2D}\cdot \epsilon_{\infty\perp}^{2D}}~,
\label{epsilon_eff}
\end{equation}
we start by considering the problem from the perspective of capacitance in anisotropic media. The capacitance $C$ of a standard parallel-plate capacitor filled with an isotropic dielectric is given by,
\begin{equation}
C = \epsilon \cdot \frac{A}{d}~,
\end{equation}
where $\epsilon$ is the dielectric constant of the material, $A$ is the area of the capacitor plates, and $d$ is the distance between the plates. However, in the case of anisotropic media, the dielectric constant is different in the parallel and perpendicular directions, so we can no longer use a single $\epsilon$ value. In an anisotropic dielectric medium, the dielectric response is characterized by $\epsilon_{\infty\parallel}^{2D}$, the dielectric constant for the in-plane (parallel) direction and $\epsilon_{\infty\perp}^{2D}$, the dielectric constant for the out-of-plane (perpendicular) direction.

Let's now model this system as having two capacitances, $C_{\parallel}^{2D}$, which is associated with the dielectric response in the parallel direction and $C_{\perp}^{2D}$, which is associated with the dielectric response in the perpendicular direction. These two components will be combined to give the overall effective dielectric response.

For capacitances in series, the effective capacitance $C^{\text{2D, eff}}$ is given by

\begin{equation}
\frac{1}{C^{\text{2D,eff}}} = \frac{1}{C_{\parallel}^{\text{2D}}} + \frac{1}{C_{\perp}^{\text{2D}}}~.
\label{cap2}
\end{equation}
Now, the expressions for capacitances  $C_{\parallel}^{\text{2D}}$ and $C_{\perp}^{\text{2D}}$  can be expressed using the dielectric constants in the corresponding directions as,
\begin{eqnarray}
C_{\parallel}^{\text{2D}} &=&\epsilon_{\infty\parallel}^{2D} \cdot \frac{A}{d}\nonumber\\
C_{\perp}^{\text{2D}} &=&\epsilon_{\infty\perp}^{2D} \cdot \frac{A}{d}~.
\label{cap3}
\end{eqnarray}
Combining the Eq.~\ref{cap2} and Eq.~\ref{cap3} into the series capacitance equation results,
\begin{equation}
C^{\text{2D,eff}} = \frac{A}{d} \cdot \frac{1}{\frac{1}{\epsilon_{\infty\parallel}^{2D}} + \frac{1}{\epsilon_{\infty\perp}^{2D}}}~.  
\label{cap3}
\end{equation}

The effective dielectric constant $C^{\text{2D,eff}}$ is also related to an effective dielectric constant $\epsilon^{\text{2D, eff}}_\infty$, which gives the overall dielectric response of the anisotropic medium, thus, 
\begin{equation}
C^{\text{2D,eff}} = \epsilon^{\text{2D, eff}}_\infty \cdot \frac{A}{d}~.
\label{cap4}
\end{equation}
Equating Eq.~\ref{cap3} and Eq.~\ref{cap4} results 
\begin{equation}
\epsilon^{\text{2D, eff}}_\infty = \Big(\frac{1}{\epsilon_{\infty\parallel}^{2D}} + \frac{1}{\epsilon_{\infty\perp}^{2D}}\Big)^{-1}
\label{cap4}
\end{equation}
Now, using harmonic and geometric mean, the Eq.~\ref{cap4} will be further used to prove Eq.~\ref{epsilon_eff}.

As a proof of concept, let's consider the harmonic mean $\mathcal{H}$ of two values $\mathcal{A}$  and $\mathcal{B}$ is given by
\begin{equation}
\mathcal{H} = \left( \frac{1}{\mathcal{A}} + \frac{1}{\mathcal{B}} \right)^{-1} = \frac{2\mathcal{A}\mathcal{B}}{\mathcal{A} + \mathcal{B}}~. 
\label{harmonic}
\end{equation}
The geometric mean $\mathcal{G}$ of the same two values is given by
\begin{equation}
\mathcal{G} = \sqrt{\mathcal{A} \cdot \mathcal{B}}~. 
\label{geometric}
\end{equation}
We want to prove that for similar values of $\mathcal{A}$ and $\mathcal{B}$, the harmonic mean approaches the geometric mean, i.e., 
\begin{equation}
\mathcal{H} \approx \mathcal{G} \quad \text{as} \quad \mathcal{A} \approx \mathcal{B}~.
\label{harmonic_geometric}
\end{equation}

let's consider $\mathcal{A}$ and $\mathcal{B}$ 
deviates from mean value $m(=\frac{\mathcal{A}+\mathcal{B}}{2}$) by $\delta_x$ and $\delta_y$ (where both $\delta_x$ and $\delta_y$ are small enough) i.e., $\mathcal{A}\approx m+\delta_x$ and $\mathcal{B}\approx m+\delta_y$. Thus, substituting in Eq.~\ref{harmonic}   
\begin{eqnarray}
\mathcal{H} &=& \frac{2(m + \delta_x)(m + \delta_y)}{(m + \delta_x) + (m + \delta_y)}\nonumber\\
&\approx&\frac{2(m^2 + m\delta_x + m\delta_y)}{2m + \delta_x + \delta_y}\approx m~.
\end{eqnarray}

Now let's expand the geometric mean in terms of $\mathcal{A}$ and $\mathcal{B}$. Substituting $\mathcal{A}\approx m+\delta_x$ and $\mathcal{B}\approx m+\delta_y$ in Eq.~\ref{geometric} results,
\begin{eqnarray}
\mathcal{G} &=& \sqrt{(m + \delta_x)(m + \delta_y)}\nonumber\\
&=&\sqrt{m^2 + m\delta_x + m\delta_y + \delta_x \delta_y}\nonumber\\
&\approx&\sqrt{m^2 + m\delta_x + m\delta_y}\nonumber\\
&\approx& m \sqrt{1 + \frac{\delta_x + \delta_y}{m}}~.
\end{eqnarray}
For small $\frac{\delta_x + \delta_y}{m}$, we can approximate the square root using a binomial expansion:
\begin{equation}
\mathcal{G} \approx m \left( 1 + \frac{\delta_x + \delta_y}{2m} \right) \approx m~.
\end{equation}

The harmonic mean $\mathcal{H}$ and the geometric mean $\mathcal{G}$ reduce to approximately $m$ when $x$ and $y$ are close. Therefore, as $x$ and $y$ approach each other, and the harmonic mean approaches the geometric mean, i.e.,
\begin{equation}
H \approx G
\end{equation}

This justifies the proof of 
Eq.~\ref{epsilon_eff} via Eq.~\ref{cap4}, i.e., 
\begin{equation}
\epsilon^{\text{2D, eff}}_\infty = \Big(\frac{1}{\epsilon_{\infty\parallel}^{2D}} + \frac{1}{\epsilon_{\infty\perp}^{2D}}\Big)^{-1} \approx \sqrt{\epsilon_{\infty\parallel}^{2D}\cdot \epsilon_{\infty\perp}^{2D}}    
\end{equation}

\onecolumngrid
\section{Details of the Material Parameters Used in This Study}
\label{Appendix-B}

\begin{longtable}{llcccccccccccccc}
    \caption{In-plane ($a$) and out-of-plane ($b$) lattice constants (in \AA) are provided. The $c$ values for the bulk structure are also given with $c = t$ (in \AA). In-plane and out-of-plane dielectric properties are included as well.} \\
    \toprule
    System & Layer & $a$&$b$&$c$ & $t$ & $\epsilon_{\infty,\parallel}^{SC}$ & $\epsilon_{\infty,\perp}^{SC}$ & $\epsilon^{2D}_{\infty\parallel}$ & $\epsilon^{2D}_{\infty\perp}$
    \\
    \midrule
    \endfirsthead
    MoS$_2$  & Bulk  &3.147 &3.147 &12.305 &$-$ &14.173 &6.241&$-$&$-$  \\
             & 1L   &3.150&3.150& 35.0&	6.062&3.272&1.164&14.117&5.409  \\
             & 2L    &3.150&3.150&40.0&12.111&5.031&1.337&14.314&6.028\\
    \midrule
    MoSe$_2$  & Bulk  &3.273 &3.273 &12.892 &$-$ &15.738 &7.995&$-$&$-$  \\
              & 1L    &3.263&3.263&35.0&6.384&3.643&1.182&15.494&6.511\\
              & 2L    &3.258&3.258&40.0&12.762&5.684&1.380&15.682&7.371\\
    \midrule
    MoTe$_2$  & Bulk  &3.494 &3.494 &13.901 &$-$ &18.978 &11.024&$-$&$-$  \\
              & 1L    &3.488&3.488&35.0&6.847&4.464&1.210&18.707&9.060  \\
              & 2L    &3.483&3.483&45.0&13.688&6.488&1.379&19.042&10.500\\
    \midrule
    WS$_2$    & Bulk  &3.147 &3.147 &12.367 &$-$ &12.955 &5.795&$-$&$-$  \\
              & 1L    &3.133&3.133&35.0&6.077&3.058&1.164&12.855&5.352\\
              & 2L    &3.128&3.128&40.0&12.156&4.646&1.335&12.998&5.783\\
    \midrule
    WSe$_2$   & Bulk  &3.282 &3.282 &12.960 &$-$ &14.499 &8.360&$-$&$-$  \\
              & 1L    &3.267&3.267&35.0&6.415&3.422&1.182&14.215&6.353& &  \\
              & 2L    &3.262&3.262&45.0&12.817&4.812&1.323&14.383&7.008\\
    \midrule
    WTe$_2$   & Bulk  &3.494 &3.494 &13.971 &$-$ &18.083 &12.616&$-$&$-$  \\
              & 1L    &3.499&3.499&35.0&6.871&4.273&1.210&17.677&8.831\\
              & 2L    &3.491&3.491&45.0&13.715&6.177&1.378&17.988&10.128\\
    \midrule
    $h$BN      & Bulk  &2.507 &2.507 &6.578 &$-$ &4.569 &2.573&$-$&$-$  \\
              & 1L    &2.494&2.494&30.0&3.317&1.394&1.068&4.569&2.363& &  \\
              & 2L    &2.492&2.492&35.0&6.634&1.673&1.125&4.551&2.440\\
    \midrule
    InSe      & Bulk  &4.034 &4.034 &16.840 &$-$ &7.457 &11.477&$-$&$-$  \\
              & 1L    &4.034&4.034&20.0&8.416&3.620&1.542&7.225&6.096\\
              & 2L    &4.043&4.043&30.0&16.808&4.550&1.915&7.336&6.801\\
    \midrule
    BP        & Bulk  &3.313 &4.496  &10.986 &$-$ &10.909 &5.843&$-$&$-$  \\
              & 1L    &3.305&4.434&20.0&5.318&5.064&1.279&16.282&5.617 \\
              & 2L    &3.305&4.434&26.0&10.637&10.77&1.536&24.881&6.827\\
    \midrule
    MoS$_2$/WS$_2$&Heterostrcuture&3.190&3.190&40.0&12.371&4.926&1.333&13.695&5.241 \\              
              \hline\hline
\end{longtable}

\twocolumngrid
\bibliography{reference.bib}

\end{document}